\documentclass[11pt,a4paper]{article}
\pdfoutput=1
\usepackage{jheppubnohead}
\usepackage[utf8]{inputenc}
\usepackage{amsmath}
\usepackage{epsfig}
\usepackage{graphicx}
\usepackage{amssymb}
\usepackage{tabularx}
\usepackage[normalem]{ulem}

\def\beqn{\begin{eqnarray}}
\def\eeqn{\end{eqnarray}} 
\def\be{\begin{equation}}
\def\ee{\end{equation}}
\def\nn{\nonumber}

\newcommand{\Xdag}{X^\dagger}
\newcommand{\Xbar}{\bar X}
\newcommand{\Xbardag}{\bar X^\dagger}
\newcommand{\psidag}{\psi^\dag}
\newcommand{\psibardag}{\bar \psi^\dagger}
\newcommand{\psibar}{\bar \psi}
\newcommand{\ubar}{\bar u}
\newcommand{\ubardag}{\bar u^\dagger}

 \title{Phenomenology of WIMPy baryogenesis models}
 
 \author[a]{Nicol\'as Bernal,}
 \author[b]{Fran\c{c}ois-Xavier Josse-Michaux}
 \author[a]{and Lorenzo Ubaldi}

 \affiliation[a]{Bethe Center for Theoretical Physics and Physikalisches Institut, \\
Universit\"at Bonn, Nu\ss allee 12, D-53115 Bonn, Germany}
 \affiliation[b]{Centro de F\'isica Te\'orica de Part\'iculas CFTP, Instituto Superior T\'ecnico, \\
Technical University of Lisbon, 1049-001 Lisboa, Portugal}

\emailAdd{nicolas@th.physik.uni-bonn.de}
\emailAdd{fxjossemichaux@gmail.com}
\emailAdd{ubaldi@th.physik.uni-bonn.de}

\abstract{
A possible connection between the abundances of baryonic and dark matter (DM) has been explored so far mostly in the context of the so-called asymmetric DM. Recently, a very different mechanism, dubbed ``WIMPy baryogenesis'', has been proposed to relate the baryon asymmetry to DM annihilation. The DM candidate is a weakly interacting massive particle (WIMP), and the usual WIMP scenario is slightly extended to accommodate baryogenesis, which is accomplished around the time of DM freeze-out.  We construct an effective field theory that encompasses a quite general class of models which implement the WIMPy baryogenesis. Under some reasonable, simplifying assumptions, we show that a good portion of the parameter space is allowed for these models, after experimental constraints are taken into account. Bounds from the LHC require that the WIMP be heavier than 400 GeV.
}

\begin{document}
\hfill {\tt Bonn-TH-2012-24}

\hfill {\tt CFTP/12-014}
\maketitle

\section{Introduction}
The presence of non-luminous and non-baryonic matter, the so-called Dark Matter (DM)~\cite{Jungman:1995df, Bergstrom:2000pn, Munoz:2003gx,
  Bertone:2004pz, Bertone:2010, Drees:2012ji, Bergstrom:2012fi}, and the existence of the baryon asymmetry of the Universe (BAU)~\cite{Tegmark:2006az,
  Komatsu:2010fb, Jarosik:2010iu} are two well established facts. Cosmic microwave background (CMB) anisotropy observations by the
Wilkinson Microwave Anisotropy Probe (WMAP) yield an accurate determination of the
total amount of baryonic matter~\cite{Komatsu:2010fb, Jarosik:2010iu},
\begin{equation}
\Omega_\text{b}h^2 = 0.02260\pm 0.00053\,,
\end{equation}
and of non-baryonic matter
\begin{equation}
\Omega_\text{DM}h^2 = 0.1123\pm 0.0035\,.
\end{equation}
The fact that the two abundances are comparable $\Omega_\text{DM}/\Omega_\text{b} \sim 5$ can be fortuitous, or may be the sign that they have a common origin.
This intriguing possibility has been vastly explored in the literature, mostly by invoking asymmetric DM (ADM) scenarios~\cite{Nussinov:1985xr,Hooper:2004dc,Farrar:2005zd,Kitano:2008tk,Kaplan:2009ag,Cohen:2009fz,Cai:2009ia,An:2009vq,Cohen:2010kn,Shelton:2010ta,Davoudiasl:2010am,Haba:2010bm,Chun:2010hz,Gu:2010ft,Blennow:2010qp,McDonald:2011zz,Allahverdi:2010rh,Dutta:2010va,Falkowski:2011xh,Heckman:2011sw,Frandsen:2011kt,Buckley:2011kk,Iminniyaz:2011yp,Cheung:2011if,MarchRussell:2011fi,Davoudiasl:2011fj,Cui:2011qe,Arina:2011cu,McDonald:2011sv,Cirelli:2011ac,Chowdhury:2011ga,Kamada:2012ht,Blum:2012nf,Tulin:2012re,Walker:2012ka,Davoudiasl:2012uw,MarchRussell:2012hi,Feng:2012jn,Ellwanger:2012yg,Okada:2012rm,Gu:2012fg}.
The common feature of all such models is that the DM abundance is determined by a matter-antimatter asymmetry in the dark sector, which in turn is connected to the baryon asymmetry in the visible sector.
The DM asymmetry can be produced prior to the BAU, can emerge from it, or can be produced during parallel and competitive processes, but in all cases the observed non-baryonic matter results from an asymmetry, implying the necessary suppression of the symmetric component.
 The ADM scenarios require a non-trivial dark sector.
 An argument in support of this choice is that the visible sector is rich and complex, why should the dark one be much simpler? 
 Although this logic is perfectly sensible, we must admit that we still know very little about the nature of DM.
 Therefore it seems reasonable to first tackle the problem keeping the dark sector as minimal as possible.
 A very simple framework is that where the DM candidate is given by just one weakly interacting massive particle (WIMP).
 Motivated by the well-known ``WIMP miracle'', one can then ask the following question: is it possible to extend minimally the WIMP scenario in order to make a connection between the DM and baryon abundances?

A few attempts in relating the relic density of such symmetric WIMP to the observed BAU have recently been made.
 The authors of~\cite{D'Eramo:2011ec} proposed a mechanism that they dubbed ``dark matter assimilation''.
 The idea is that singlet DM, $\chi$, (they study a bino-like neutralino as an example) is efficiently depleted in the early Universe by being assimilated into new quasi-stable heavy states, $\Psi$ and $\tilde\Psi$ via the reaction $\chi \Psi \to \tilde\Psi \phi$, where $\phi$ is a Standard Model (SM) particle. These new heavy states carry the baryon asymmetry.
 Their subsequent annihilations and decays (into DM and SM quarks) yield the correct DM relic density and BAU. The minimal version of such models is quite economical as it requires only the addition of singlet DM and two new heavy states on top of the SM particle content. 
Another mechanism, that goes under the name of  ``baryomorphosis'' and shares some similarities with the DM assimilation, was studied in~\cite{McDonald:2011zz, McDonald:2011sv}. In the baryomorphosis scenario a large asymmetry is originally stored in a heavy scalar field, which then decays into two colored scalars, the annihilons, with mass at the TeV scale. Their late annihilations into singlet scalar DM particles set the correct DM abundance and baryon asymmetry.

More recently, Cui, Randall and Shuve introduced a new mechanism, which they refer to as ``WIMPy baryogenesis''~\cite{Cui:2011ab}.
 Here the DM is a Dirac fermion\footnote{This is the possibility studied in their explicit models.
 A Majorana fermion or a scalar can in principle work as well.} with a mass of at least a few hundreds GeV.
 One also needs to add two new states, one heavy ($\sim$ TeV) charged under the SM quantum numbers, $\psi$, the other one very light and uncharged, $n$.
 DM annihilates into $\psi$ and a quark (or a lepton), and $\psi$ subsequently decays into the light, sterile state $n$, storing the negative asymmetry in baryon (lepton) number in a sequestered sector.
 The DM relic density is that of a thermal WIMP, and by the time the DM annihilations freeze out one has generated a baryon asymmetry, either directly, when the annihilation is into $\psi$ plus quark, or via leptogenesis, when the annihilation is into $\psi$ plus lepton.
 In the latter case the lepton asymmetry has to be generated before the electroweak phase transition (EWPT) so that sphaleron processes can transfer the asymmetry to the baryon sector, with the consequence that the masses of DM and $\psi$ must be at least $\mathcal{O}$(TeV).
 In the case where DM annihilates into quarks, baryogenesis can occur after the EWPT and masses down to a few hundreds GeV are allowed. 

There is a conceptual difference between the WIMPy baryogenesis framework and the typical ADM frameworks.
Whereas both aim at predicting the correct baryon and DM abundances, they stem from different motivations.
ADM models are built in order to explain the observed ratio $\Omega_\text{DM}/\Omega_\text{b} \sim 5$, understood as non-fortuitous.
  WIMPy baryogenesis models, instead, are based on taking seriously another coincidence: the WIMP miracle.
Adding only a few new ingredients, one keeps such a miracle and can easily accommodate baryogenesis. On the one hand, ADM gives up the WIMP miracle; on the other hand, WIMPy baryogenesis does not explain why $\Omega_\text{DM}/\Omega_\text{b} \sim 5$, but as we show in this work, for values of the parameters that we consider natural, and are still well within experimental constraints, these models give the right numbers for the DM abundance and the BAU. 

The aim of the current paper is to study a general class of models that implements the WIMPy baryogenesis mechanism.
 In the spirit of keeping the models as simple as possible we identify the minimal particle content that does the job.
 In Ref.~\cite{Cui:2011ab}, the possibility of DM annihilations generating a lepton asymmetry, which is then converted into a baryon asymmetry through sphalerons, was somewhat emphasized.
 In this work we only consider models where DM annihilates into a quark plus an exotic, heavy antiquark, thus producing the BAU directly.
 We find the latter scenario more appealing than the former for a few reasons.
 First, for the sake of simplicity, we can avoid the extra step with sphalerons needed in the leptogenesis case.
 Second, because the exotic quark is colored, as the name suggests, the LHC has a chance of discovering it or, alternatively, it could put some severe bounds on the model.
Third, given that the DM  directly couples to quarks, we can in principle hope for a direct detection signal, which would certainly not be there if DM interacted only with leptons. Thus, the scenario we study here seems more testable than the WIMPy leptogenesis, although the latter still remains a possibility worth exploring.

Our approach is a little different than the one in~\cite{Cui:2011ab}.
They build a UV-complete model, including pseudoscalars that mediate the DM annihilation and can have masses of the same order as the DM and the exotic quark; the Lagrangian they use as the basis for their calculations only includes renormalizable interactions. 
We do not include any mediators in our model instead, assuming that they are all much heavier than the fermions, and in the spirit of an effective field theory (EFT) we write only four-fermion interaction terms.
Doing so we have a more generic and richer class of models that allows for new operators and for new DM annihilation channels.

The paper is organized as follows. In Section~\ref{EFT} we present the particle content and the Lagrangian of the model, then we discuss how the baryon asymmetry is generated. In Sections~\ref{sec:collider}, \ref{sec:cosmology} and \ref{sec:direct} we study the experimental constraints on the model from the LHC, cosmology, and DM direct detection. We summarize our conclusion in Section~\ref{sec:summary}. Some technical details are included in the final appendices.



\section{An effective-field-theory approach}\label{EFT}


\subsection{Field content of the model and Lagrangian}\label{EFT1}
We add to the SM the minimal particle content which is needed in order to have a successful WIMPy baryogenesis.
All the new particles are fermions.
We consider vector-like gauge singlet DM $X$ and $\Xbar$, vector-like exotic quark color triplets $\psi$ and $\psibar$, and a massless singlet $n$, into which the exotic quark decays, as we will explain in the next subsection. 

\begin{table}[!htb]
\centering
\begin{tabular}{ | c | c | c | c | c | c |}
\cline{2-6}
\multicolumn{1}{c|}{} & $\boldsymbol{SU(3)_c}$ & $\boldsymbol{SU(2)_L}$ & $\boldsymbol{Q_{U(1)_y}}$ & $\boldsymbol{Q_{U(1)_B}}$ & $\boldsymbol{\mathbb{Z}_4}$ \\
\hline
$X$               & 1 & 1 & 0 & $0$ & $+i$ \\
$\Xbar$           & 1 & 1 & 0 & $0$ & $-i$ \\
$\psi$     & 3 & 1 & $+2/3$ & $+1/3$ & $+1$ \\
$\psibar$ & $\bar 3$ & 1 & $-2/3$ & $-1/3$ & $+1$ \\
$n$       & 1 & 1 & 0 & 0 or $+1$ & $+1$ \\
$\ubar$   & $\bar 3$ & 1 & $-2/3$ & $-1/3$ & $-1$ \\
$\bar d$          & $\bar 3$ & 1 & $+1/3$ & $-1/3$ & $-1$\\
\hline 
\end{tabular}
\caption{\sl \textbf{\textit{Particle content of the model.}} $\ubar$ and $\bar d$ are the right-handed up and down quarks of the SM. The rest of the SM quarks also have charge $-1$, while all the leptons are neutral under the $\mathbb{Z}_4$ symmetry. The reason for these charge assignments is explained in Appendix~\ref{app:discrete}.  \label{tab:content}}
\end{table}

\paragraph{Notation and conventions}\label{sec:notation}
We use the two-component spinor formalism and we follow closely the conventions of Ref.~\cite{Dreiner:2008tw}. The advantage of such a formalism versus the four-component spinor one is that Fierz identities are easier, which greatly simplifies the task of finding a complete, linearly independent basis of dimension six operators.
Occasionally, it proves convenient to switch back to four-component notation, in which case we denote the spinors as follows
\be \label{eq:4cpt}
\chi = 
\begin{pmatrix}
X \\
\Xbardag
\end{pmatrix} \quad
\Psi = 
\begin{pmatrix}
\psi \\
\psibardag
\end{pmatrix} \quad
P_R U = 
\begin{pmatrix}
0 \\
\ubardag
\end{pmatrix}.
\ee
From eq.~\eqref{eq:4cpt} it is evident that $X$ and $\Xbardag$ represent DM particles, while $\Xbar$ and $\Xdag$ anti-DM particles. \newline

A discrete symmetry is needed in our model, in order to stabilize the DM and to avoid dangerous decays of the exotic quark that could spoil the baryon asymmetry.
Note that all the SM quarks carry charge $-1$ under the $\mathbb{Z}_4$, while all the leptons and the Higgs boson are neutral, so that the familiar renormalizable SM Lagrangian is unchanged.
The $\mathbb{Z}_4$ also guarantees the stability of the proton that, having charge $(-1)^3=-1$, can never decay into lighter mesons and leptons, which are neutral.
Neutron-antineutron oscillations can in principle occur, given that we have baryon number violation, but the bounds do not pose a real challenge to these models, as we explain in Appendix~\ref{app:nn}. 

The reason for the choice of the discrete charges assigned to the various fields is discussed in Appendix~\ref{app:discrete}. In the same Appendix we also show why, restricting ourselves to global discrete Abelian groups, the minimal choice within our field content is $\mathbb{Z}_4$.
We emphasize the fact that the $\mathbb{Z}_4$ we impose is not generic for the WIMPy baryogenesis mechanism. Rather, it is tied to our specific models.
 It is possible that other models with different symmetry groups do the job, although we have not investigated such a possibility. 

In order to study the phenomenological implications of this model, we write down an effective Lagrangian that includes {\em all} the dimension six operators $\mathcal{O}_i$ (four-fermion operators) consistent with the field content and the quantum numbers listed in Table~\ref{tab:content},
\be \label{eq:effL}
\mathcal{L} \supset \frac{1}{\Lambda^2} \sum_i \lambda_i^2 \mathcal{O}_i\,.
\ee
Here we have chosen to parametrize the couplings as $(\lambda_i / \Lambda)^2$, with dimensionless $\lambda_i$'s and a fixed mass scale $\Lambda$. 
The list of 20 (plus Hermitian conjugates) operators $\mathcal{O}_i$ is given in Appendix~\ref{app:list}. 
Given the particle content we consider, this set of operators spans a complete and irreducible basis. 

Note that we {\em chose} the basis with only scalar operators.
However, this does not mean that a UV-complete theory where the exchanged particle is a vector, for instance, cannot be mapped into our EFT.
As an example, consider a lagrangian that includes the terms $g\,A_\mu\,(\Xbardag\, \bar\sigma^\mu\,X +\Xdag\, \bar \sigma^\mu\, \Xbar) + g'\, A_\mu\, \ubar\, \sigma^\mu\, \psibardag$, relevant for DM annihilation.
If we integrated out $A_\mu$ we would get the dimension 6 operators $\frac{g\,g'}{M_A^2} (\Xbardag \bar \sigma^\mu X + \Xdag \bar \sigma^\mu \Xbar)  (\ubar\, \sigma_\mu\, \psibardag)$, which are equivalent to $2\frac{g\,g'}{M_A^2} [(\Xbardag \psibardag)(X\ubar)+(\Xdag \psibardag)(\Xbar \ubar)]$, by using a Fierz identity.
These latter operators are included in our basis.
This also shows that a vector exchange in the $s$-channel is equivalent to a scalar exchange in the $t$-channel for this DM annihilation. 
In general, all the possible vector- (e.g. $(\Xbardag \bar \sigma^\mu X)(\ubar  \sigma_\mu \psibardag )$) and tensor- (e.g. $(X \sigma^{\mu\nu} X)(\psi  \sigma_{\mu\nu} \ubar)$) operators can be related to the scalar ones in our list.
The procedure to map a UV-complete theory into our EFT is the one outlined in the example just given: First, integrate out the heavy exchanged particles to obtain four-fermion operators, then use Fierz identities to reduce them to scalars.

The EFT approach is valid only as long as the biggest momentum involved in the processes we are considering, $k_\text{max}$, is such that $k_\text{max} \frac{\lambda_i}{\Lambda} < 1$.
In our study, $k_\text{max}$ is given by some temperature, $T$, before the DM freezes out.
As we will explain in the next subsection, the baryon asymmetry should build up between the time the washout processes freeze out, $z=z_\text{WO}$, and the time of DM annihilation freeze-out, $z=z_\text{FO}$, where $z\equiv m_\chi / T$.
 As in usual thermal WIMP scenarios, $z_\text{FO}\sim 25$, while we find that typically $z_\text{WO} \sim 10$.
 Given that we scan over a DM mass range up to 1 TeV, with $z_\text{WO} = 10$ giving the reference value for the highest temperature, we have $k_\text{max} \sim m_\chi / z_\text{WO} \sim 100$ GeV.
The condition for the validity of the EFT approach then translates into a bound for the couplings, namely
\begin{equation}\label{bound}
\frac{\lambda_i}{\Lambda} < (100 \ {\rm GeV})^{-1}\,.
\end{equation}
In the following numerical evaluations we fix $\Lambda = 10$ TeV, which in turn implies $\lambda_i < 100$~\footnote{
This last bound can lead to some confusion, thus a comment is in due order.
 In the parametrization we {\em chose} for the Lagrangian~\eqref{eq:effL}, the couplings $\lambda_i$ can be thought of as dimensionless Yukawa's.
 In a given UV-completion of our theory, they would be subject to the usual perturbative bound, $\lambda_i < 4 \pi$, which seems to be in conflict with our $\lambda_i < 100$.
 The point is that the only sensible condition for our EFT is expressed as $\frac{\lambda_i}{\Lambda} < (100 \ {\rm GeV})^{-1}$.
 One could {\em always} keep $\lambda_i$ below $4 \pi$ by lowering the scale $\Lambda$.
 In other words, the bound $\lambda_i \lesssim 100$ is just a consequence of our parametrization and the choice of fixing $\Lambda$ to 10 TeV.
 Values of $\lambda_i$ bigger than $4 \pi$ can just be thought of as lowering the scale $\Lambda$ below 10 TeV. Different parametrization would obviously lead to the same physical results. For example, other DM studies based on EFT approach (e.g.~\cite{Cao:2009uw,Zheng:2010js,Cheung:2012gi}) assume instead $\mathcal{O}(1)$ coefficients and constrain one scale $\Lambda_i$ per operator.}.

To keep the number of parameters in the numerical analysis manageable, we set some equalities among the relevant couplings and we relabel them for the ease of the discussion\footnote{In Appendix~\ref{app:list} the reader can find the operator $\mathcal{O}_i$  associated to $\lambda_i$, with $i=1\dots 20$.}:
\begin{itemize}
\item $\lambda_{s1}  \equiv   \lambda_1=\lambda_3$, \quad $\lambda_{s2}  \equiv   \lambda_2=\lambda_4$: couplings for DM (or anti-DM) annihilation into $ \psi \ubar$ and $\psidag \ubardag$ in the $s$-channel;  
\item $\lambda_t \equiv   \lambda_5=\lambda_6$: couplings for DM (or anti-DM) annihilation into $ \psi \ubar$ and $\psidag \ubardag$ in the $t$-channel; 
\item $\lambda_\text{WO} \equiv   \lambda_9=\lambda_{11}$: couplings responsible for the ``pure" washout processes\footnote{The ``pure'' and ``mixed'' washout processes are defined in the next subsection.};
\item $\lambda_n \equiv   \lambda_{20}$: coupling implying the decay $\psi \to \bar d \bar d n$. 
\end{itemize}
The couplings $\lambda_{s1}$, $\lambda_{s2}$ and $\lambda_t$ are also involved in the ``mixed'' washout processes, that are related to the annihilations by crossing symmetry.

The relations among the couplings listed above are imposed by hand at the level of the EFT, but we can think of them as inspired by possible UV completions of the model.
For example, in the toy model we mentioned earlier, with the heavy vector $A_\mu$, we would have in the EFT the relation $\frac{\lambda_t^2}{\Lambda^2} \equiv \frac{\lambda_5^2}{\Lambda^2} = \frac{\lambda_6^2}{\Lambda^2} = \frac{g\,g'}{M_A^2}$.
As another example, in a model where the annihilation proceeds through a scalar exchange in the $s$-channel, like in~\cite{Cui:2011ab}, we would obtain relations similar to the one we wrote for $\lambda_{s1}$ and $\lambda_{s2}$.
The reason for keeping two couplings in the $s$-channel is evident if we write the corresponding operators in four-component notation
\be \label{eq:4s}
\frac{1}{\Lambda^2}\left[(\lambda_{s1}^2 + \lambda_{s2}^2) (\bar \chi \chi^c + \bar \chi^c \chi)   + (\lambda_{s1}^2 - \lambda_{s2}^2) (\bar \chi \gamma_5 \chi^c - \bar \chi^c \gamma_5 \chi)   \right] \left[\bar U P_L \Psi\right] \, .
\ee 
This expression shows that, for general $\lambda_{s1}$ and $\lambda_{s2}$, we have a mixture of scalar and pseudoscalar channels.
 In our analysis we will consider two limiting cases:
\beqn
\lambda_s^2 & \equiv  \lambda_{s1}^2 = +\lambda_{s2}^2  & \qquad \text{scalar }s\text{-channel}\label{sscalar}, \\
\lambda_p^2 & \equiv  \lambda_{s1}^2 = -\lambda_{s2}^2  & \qquad \text{pseudoscalar }s\text{-channel}\label{spscalar}.
\eeqn

Many couplings do not appear in the relabeling list above. Two of them, $\lambda_7$ and $\lambda_8$, contribute to DM annihilation into two quarks and to DM direct detection, and will be discussed in Section~\ref{sec:direct}. 
The couplings for the annihilation $X \Xbar \to \psi \psibar$
\be
\lambda_{\psi\psi}  \equiv  \lambda_{13}=\lambda_{14}=\lambda_{15}=\lambda_{16}=\lambda_{17}=\lambda_{18}=\lambda_{19} 
\ee
deserve some discussion.
 This channel,  corresponding to a CP-conserving DM annihilation, is kinematically allowed only when $m_\psi < m_\chi$.
In our numerical analysis that follows, we scan for $m_\psi$ in the range $[0.8,\,2]\times m_\chi$.
The coupling $\lambda_{\psi\psi}$ is completely irrelevant in most of this range, namely between $m_\chi$ and $2\,m_\chi$, because this channel is not open.
It would play a role in the rest of the range, $m_\psi < m_\chi$, where this process would compete against the CP-violating annihilations that involve the couplings $\lambda_{s,\,p,\,t}$.
The former has to be suppressed compared to the latter to generate a sizable baryon asymmetry.
One has then two choices: to keep the parameter $\lambda_{\psi\psi}$ in the analysis, or to drop it under the assumption that, for $m_\psi < m_\chi$, it must be much smaller than $\lambda_{s,\,p,\,t}$ in order not to spoil the BAU.
We decided to opt for the second one, given that only a small part of the parameter space is affected anyway, for the sake of keeping the number of parameters minimal.
 
The remaining two couplings, $\lambda_{10}$ and $\lambda_{12}$, are included in the list in Appendix~\ref{app:list} for completeness, but do not play any role in the rest of our discussion.

\subsection{Generation of the baryon asymmetry via WIMP annihilations}\label{sec:sub2}

DM annihilates into a $\bar u$ quark and an exotic quark, $\psi$. 
We emphasize that in these models the DM, despite being a Dirac fermion, annihilates with itself rather than with its antiparticle.
 This can be seen explicitly from eq.~\eqref{eq:4s}, for example.
During these annihilations, CP is violated and an $u$-quark number asymmetry is created, in an equal and opposite amount to a $\psi$-number asymmetry\footnote{In principle, one could worry about sphaleron processes even in this context. If the BAU generation occurs before EWPT, the baryon asymmetry is converted into a conserved $B$-$L$ charge, while the $B$ charge is set to zero by sphalerons.
After their freeze-out, the $B$-$L$ number is split into $B$ and $L$ number.
However, given the masses considered in this work, the baryon asymmetry generation occurs after EWPT, with no reprocessing effects from sphalerons.}.  
After being produced, the exotic quark decays into a SM singlet fermion, $n$, plus two antiquarks: $\psi \to \bar d \bar d n$. It is crucial that $\psi$ does not decay only into SM quarks, since that could eliminate the asymmetry.
 The $\mathbb{Z}_4$ symmetry prevents such dangerous decays.
To avoid overclosing the Universe, $n$ has to be light\footnote{An interesting possibility is that $n$ constitutes the extra degree of radiation at BBN. We will explore this in more detail in future work.}. For the sake of simplicity we take it to be massless. Two scenarios are contemplated:
\begin{enumerate}
\item The singlet $n$ carries baryon number $+1$.
 In this case the decay $\psi \to \bar d \bar d n$ is baryon-number-conserving, but $n$ is sequestered in a sterile sector, so we are left with a net baryon number in the visible sector.
 The overall process violates the SM baryon number.
\item The singlet $n$ does not carry baryon number.
 The decay of $\psi$ explicitly violates baryon number and it contributes to the baryon asymmetry.
\end{enumerate}
Both cases satisfy the first Sakharov condition~\cite{Sakharov:1967dj}, that is {\em baryon number violation}. The other two conditions are also easily satisfied: {\em CP violation} is achieved with complex couplings $\lambda_i$ and with the interference between tree-level and one-loop diagrams (see Appendix~\ref{app:epsilon}); {\em departure from thermal equilibrium} is automatically implemented, given that WIMP annihilation around freeze-out is out of equilibrium.

\begin{figure}[!h]
\begin{center}
\includegraphics[width=0.7\textwidth]{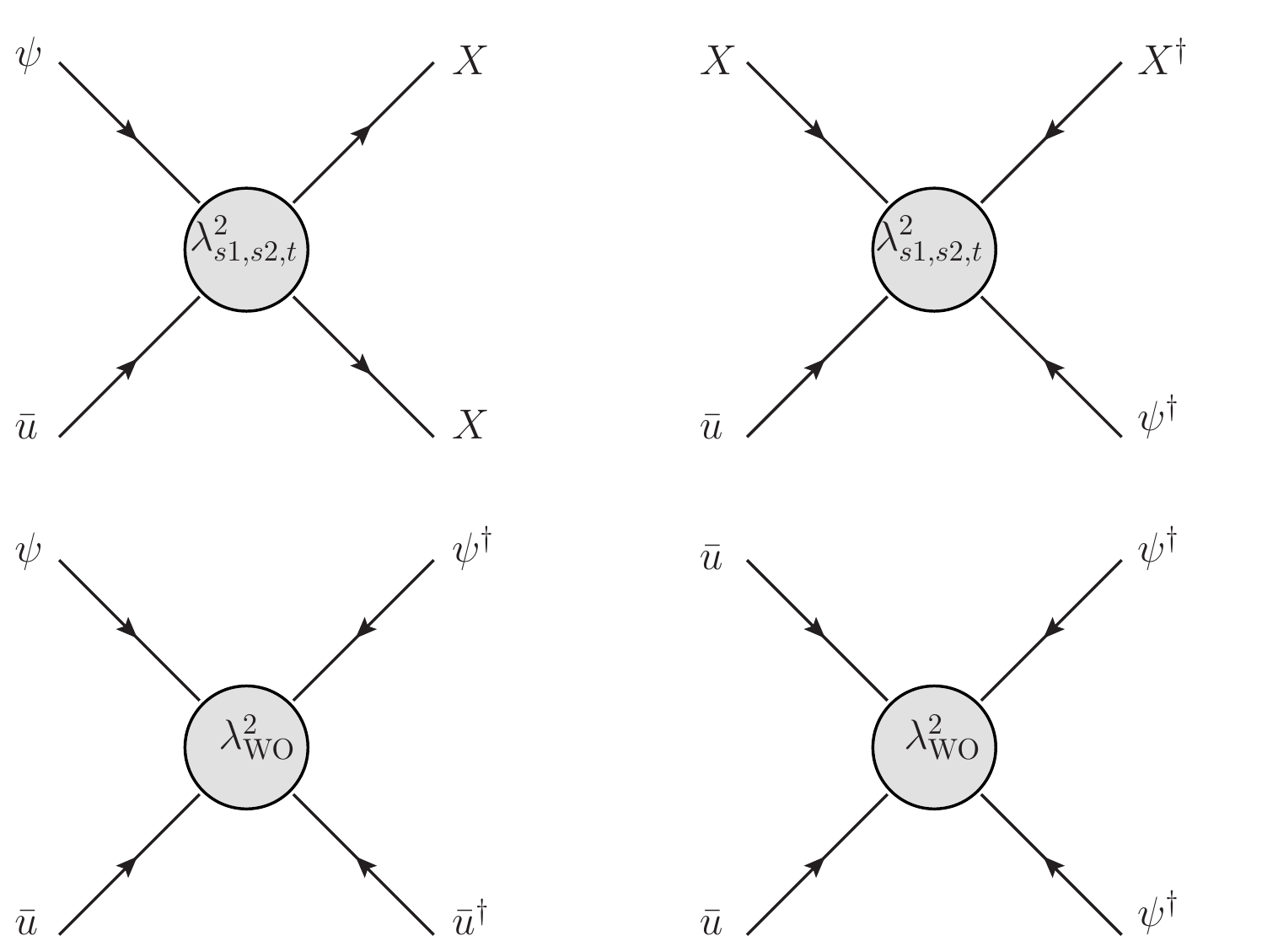}
\end{center}
\vspace{-0.8cm}
\caption{\sl \textbf{\textit{Washout processes.}} We show mixed washout processes in the upper two diagrams and pure washout processes in the lower two.}
\label{Fig:washout}
\end{figure}

There are processes that can potentially wash out the asymmetry. They are shown schematically in figure~\ref{Fig:washout}.
The processes $\psi \ubar \to XX$ and $\ubar X \to \Xdag \psidag$, involving DM particles, are referred to as ``mixed'' washout.
They are obtained from the DM annihilation diagrams by crossing symmetry, and so involve the same couplings $\lambda_{s1}$, $\lambda_{s2}$, $\lambda_t$.
The processes in the two lower diagrams, that involve only $\psi$ and $\ubar$, go under the name of ``pure'' washout.
We will discuss further about these processes in Section~\ref{sec:cosmology}.
 
A very important result, emphasized by the authors of \cite{Cui:2011ab}, is that {\em in order to produce a significant baryon asymmetry, washout processes must freeze out before WIMP freeze-out}. To achieve this early washout freeze-out, one needs either a $\psi$ heavier than the DM, $m_\psi > m_\chi$, so that the washout is Boltzmann suppressed while DM is still annihilating, or a small couplings, such that the washout cross section is small compared to the annihilation cross section.


\section{Constraints from the LHC} \label{sec:collider}

One of the new particles that we need in our models, $\psi$, is colored, which makes it a good candidate to be discovered at the LHC. Alternatively, the LHC can put severe bounds on these models.
$\psi$ can be pair-produced at the hadron collider, through the process $q \bar q \to g \to \psi \bar \psi$, where a quark and an antiquark annihilate into a gluon, which then splits into $\psi$ and $\psibar$.
 Given that each $\psi$ decays into two (anti-)down quarks plus a singlet $n$, the signature to look for is four jets plus missing energy.
 Both the CMS~\cite{CMS-PAS-SUS-11-016} and the ATLAS~\cite{ATLAS} collaborations search for such a signature in the context of supersymmetry (SUSY), and they put bounds on the masses of gluinos and squarks from the process $pp \to \tilde g \tilde g \to 4 j + {\not} E_T$, where the missing energy is carried away by the lightest neutralinos.
 In the case of a massless neutralino, which is considered in~\cite{ATLAS}, the SUSY process is completely analogous to the one we are interested in, with $\psi$ in place of the gluino.
 The production cross section differs only by a group theory factor, since $\psi$ is a color triplet, while the gluino is an octet.
 Once that is taken into account, we can translate the bound\footnote{This is the bound from~\cite{ATLAS} when the squark is very heavy, which is appropriate in our EFT context, where all the mediators are very heavy and have been integrated out.} on the gluinos, $m_{\tilde g} \gtrsim 960$ GeV, into
\be \label{eq:LHCbound}
m_\psi \gtrsim 800 \ {\rm GeV}.
\ee

In these models DM has to annihilate into a quark, whose mass can be neglected, and an exotic antiquark, $\psi$, in order to generate a baryon asymmetry. Therefore we have the kinematical bound
\be \label{eq:kinbound}
m_\psi < 2\,m_\chi.
\ee
Thus, the DM has to be heavier than $\sim 400$~GeV.


\section{Constraints from cosmology} \label{sec:cosmology}
In this section we assess the impact of the DM relic density and BAU constraints on our parameter space. To that end, we make use of Boltzmann equations, detailed in Appendix~\ref{app:BEs}, for the evolution of the DM density and the baryon asymmetry.

In the numerical evaluations, we fix $\Lambda=10$~TeV, and we scan over $m_\chi$ and $m_\psi$, with $m_\chi \leq 1$ TeV, and  $800$ GeV $\leq m_\psi \leq 2 \ m_\chi$. The lower and upper bounds on $m_\psi$ are dictated by the LHC and the kinematics of DM annihilations, respectively.
We recall that the dimensionless couplings $\lambda_i$'s have an upper bound of $\sim 100$, as required by the validity of the EFT approach.

\subsection{Dark matter relic abundance}

The Boltzmann equation governing DM annihilations is given in eq.~\eqref{eqDM}. As we work in a first order expansion in the asymmetries, the small contributions proportional to e.g. $\epsilon\times Y_{\Delta u}$ are neglected, and the  DM relic density is set by the typical thermal freeze-out.
In principle the annihilations can have a CP-conserving (CPC) and a CP-violating (CPV) contribution. The CPC channels, governed by the couplings $\lambda_{\psi\psi}$, $\lambda_7$ and $\lambda_8$ reduce the WIMPy baryogenesis efficiency, so they have to be subdominant if one wants to generate a sizable baryon asymmetry. On the one hand, $\lambda_7$ and $\lambda_8$ can be somewhat suppressed to avoid conflict with direct detection searches, as we explain in section~\ref{sec:direct}. On the other hand, we have to assume $\lambda_{\psi\psi}\ll \lambda_{s1,2}$,~$\lambda_{t}$ when this channel is kinematically open, i.e. for $m_\psi / m_\chi \leq 1$.

\begin{figure}[!t]
\begin{center}
\includegraphics[width=10cm]{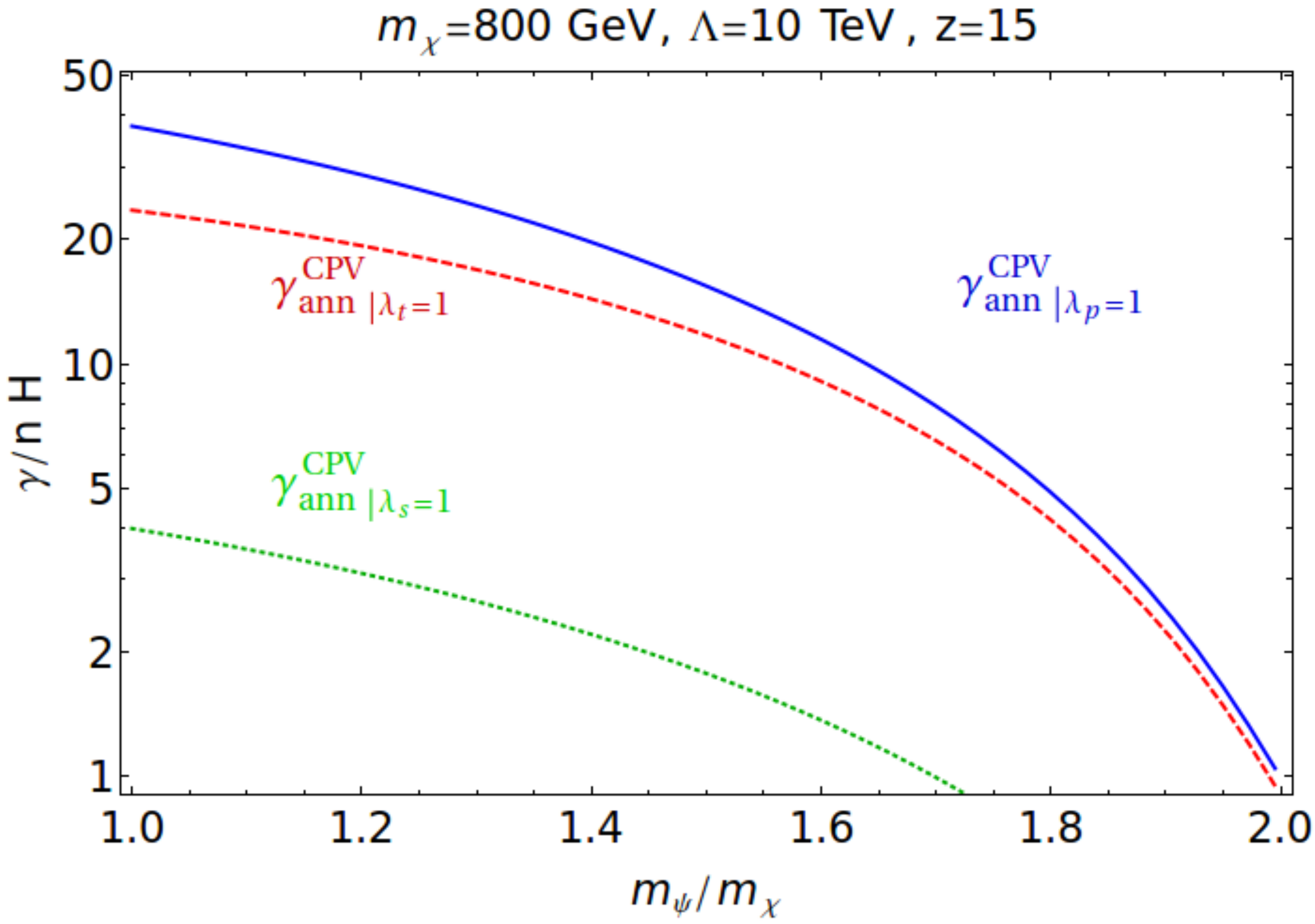}
\end{center}
\vspace{-0.8cm}
\caption{\sl \textbf{\textit{Dark Matter annihilation rate as a function of the mass ratio $\boldsymbol{m_\psi /m_\chi}$, for various limit cases.}}
From top to bottom, the curves represent the pseudoscalar $s$-channel, the $t$-channel and the scalar $s$-channel. For illustration, all respective couplings have been set to $1$.}
\label{Fig:Rates}
\end{figure}

It is  instructive to inspect the CPV DM annihilation rate\footnote{See Appendix~\ref{app:BEs} for the definition of the rate.}.
In the low temperature limit (large $z \equiv m_\chi / T$) it reads
\begin{equation} \label{eq:DMrate}
\gamma_\text{ann}^\text{CPV}\simeq \frac{m_\chi^8}{64\,\pi^4\Lambda^4}\frac{e^{-2z}}{z^3}\,\left(1-x^2\right)^2 \left[2\lambda_p^4-2\lambda_p^2\,\lambda_t^2\,x+\left(1+x^2\right)\lambda_t^4+\frac{3}{z}\lambda_s^4\right]\,,
\end{equation}
where $x\equiv m_\psi/(2\,m_\chi)$. 
Eq.~\eqref{eq:DMrate} shows the known result that the scalar $s$-channel is velocity suppressed compared to the pseudoscalar one.
Since in the rest of this section we will analyze the scalar, pseudoscalar, and $t$-channel cases
 separately, it is worth having in mind their relative contributions. 
 To this end, we  display in figure~\ref{Fig:Rates} the rate $\gamma_\text{ann}^\text{CPV}$ (normalized to $n_\chi^{eq}(z)H(z)$) for  various limiting cases. It is clear that the pseudoscalar and $t$ channels  dominate. Thus, a larger $\lambda_s$, compared to $\lambda_p$ and $\lambda_t$, will be needed to get the right relic density.
  The kinematic suppression    of the rates is also manifest in figure~\ref{Fig:Rates} for $m_\psi/m_\chi\gtrsim 1.8$: the larger the ratio
    $m_\psi /m_\chi$ is, the larger the couplings will have to be in order to compensate for such a suppression.

\begin{figure}[!h]
\begin{center}
\includegraphics[width=6cm]{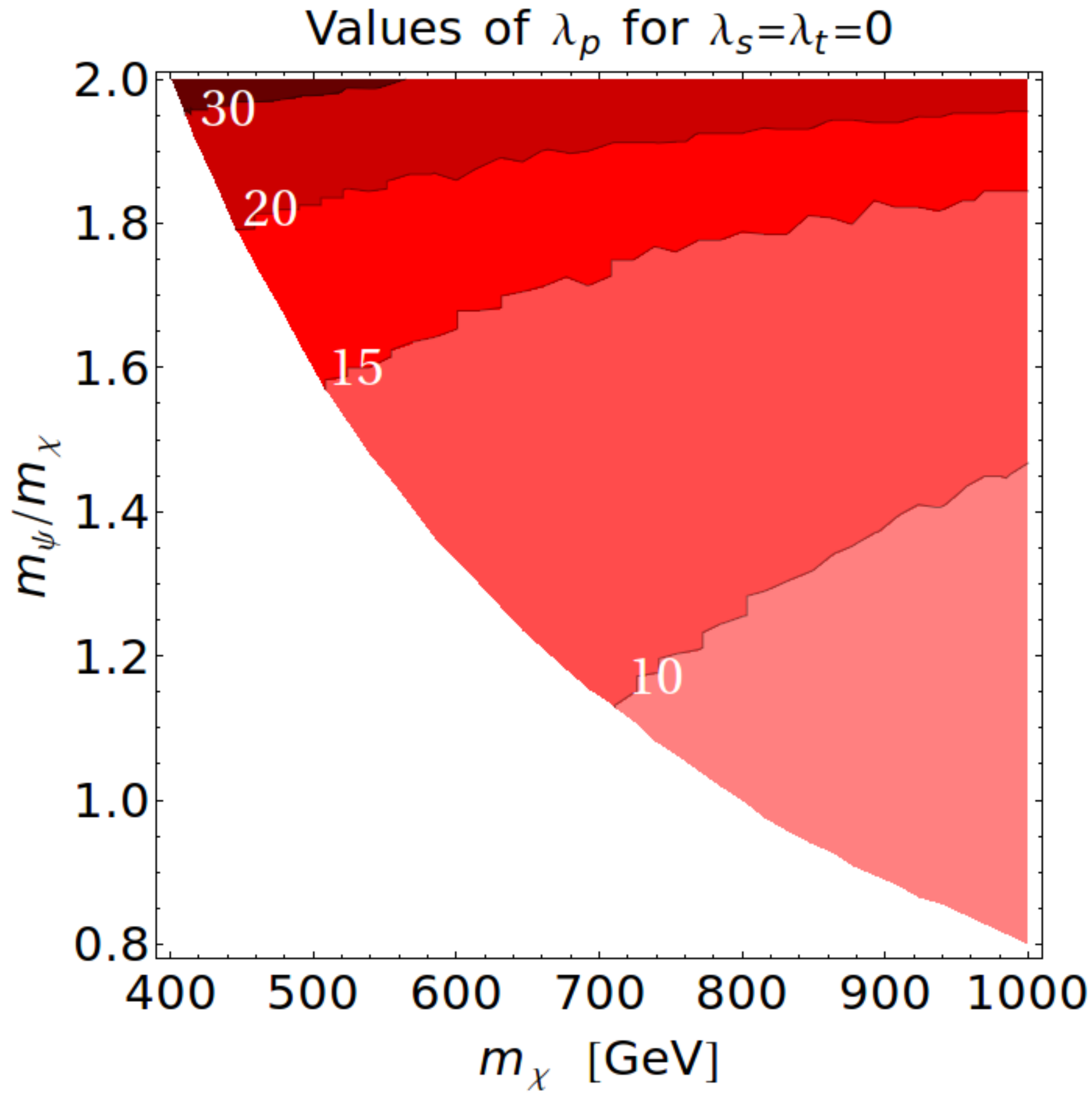}\qquad\includegraphics[width=6cm]{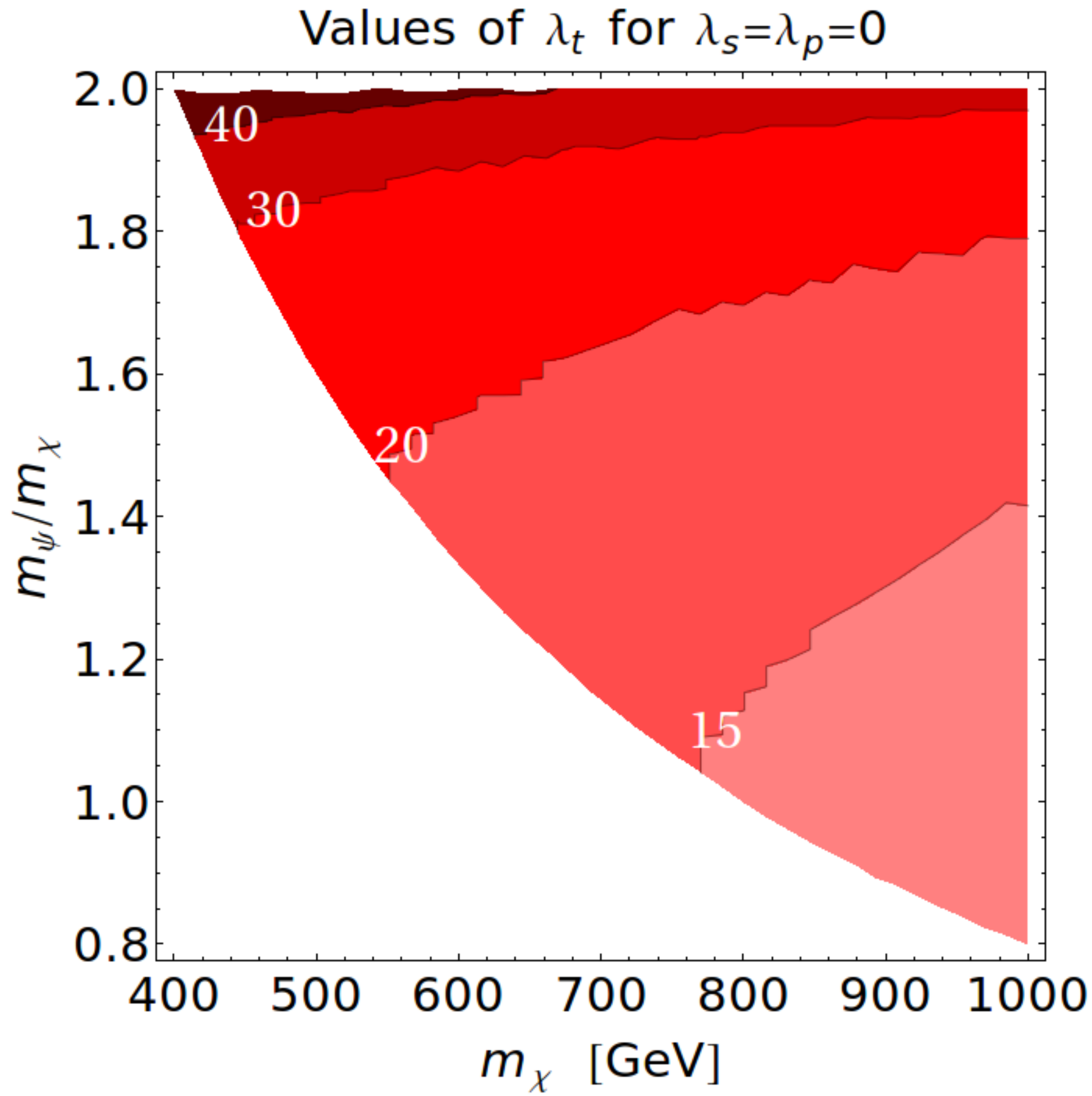}\\
\includegraphics[width=6cm]{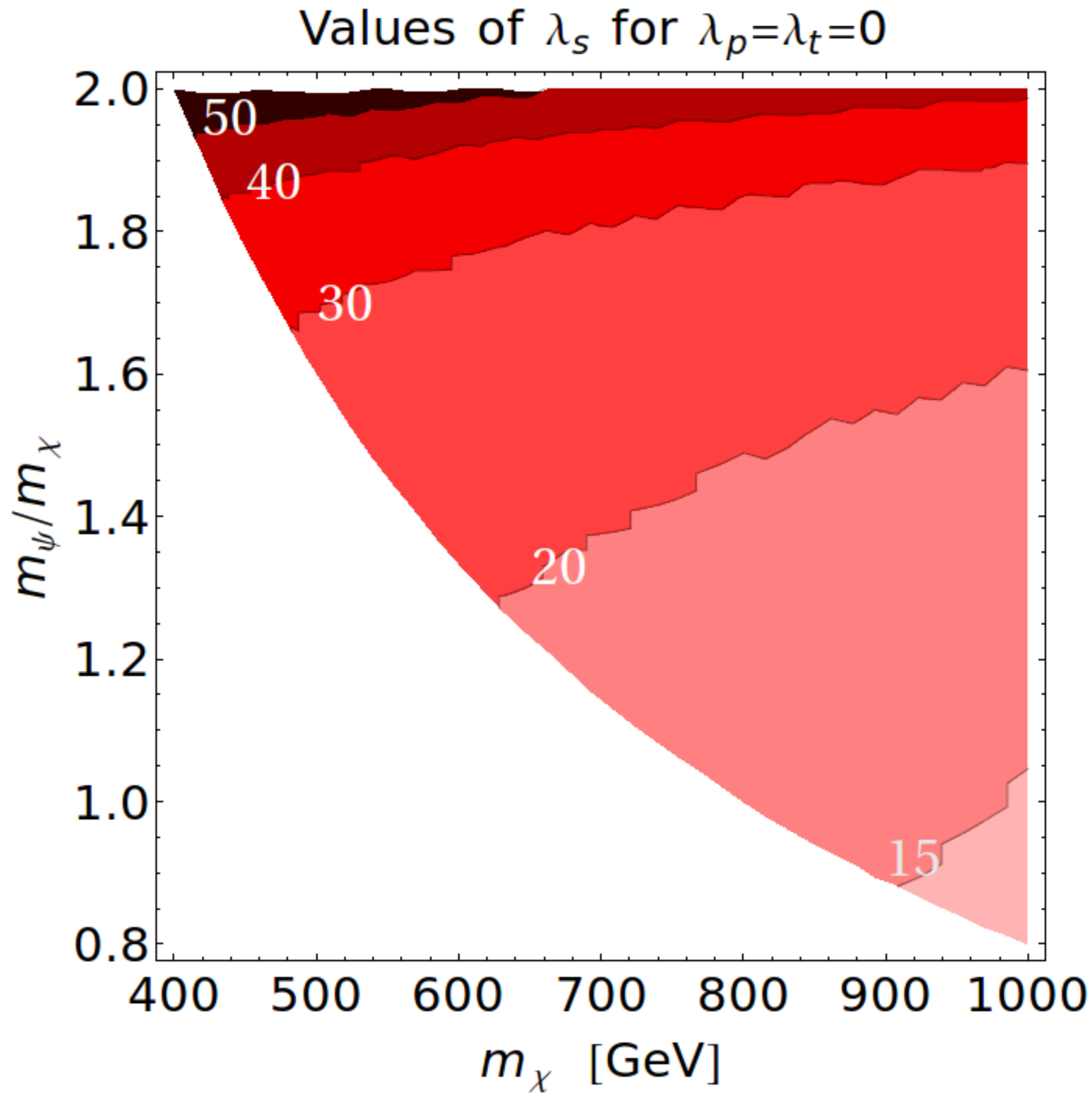}
\end{center}
\vspace{-0.8cm}
\caption{\sl \textbf{\textit{Contour level for the couplings $\boldsymbol{\lambda_p}$ (upper left pane), $\boldsymbol{\lambda_t}$ (upper right pane) and $\boldsymbol{\lambda_s}$ (lower pane) needed for generating the DM relic density in the $\boldsymbol{[m_\psi/m_\chi,\,m_\chi]}$ plane.}}
The white lower left part corresponds to $m_\psi \leq 800$ GeV, which is excluded by the LHC.
}
\label{Fig:GrapheLsLpLt}
\end{figure}

The expected constraints are confirmed in figure~\ref{Fig:GrapheLsLpLt}, which shows contour levels for the coupling $\lambda_p$ (upper left pane), $\lambda_t$ (upper right pane) and $\lambda_s$ (lower pane) needed for generating the DM relic density abundance measured by WMAP, in the $[m_\psi/m_\chi , \,m_\chi ]$ plane.
 The white lower left region in each plot corresponds to a $\psi$ mass already ruled out by the LHC ($m_\psi<800$~GeV).
In general, the couplings $\lambda_{p,\,s,\,t}$ have to be larger than 10.
We see that for a fixed $m_\psi/m_\chi$ ratio, smaller couplings are needed for larger $m_\chi$. Indeed, with the increase of $m_\chi$ the DM relic abundance gets reduced due to the thermal average (for a fixed DM annihilation cross section), and to the increase of the DM cross section itself, hence smaller couplings are required. Oppositely, for a fixed DM mass, as shown in figure \ref{Fig:Rates}, the annihilation cross section decreases for larger $\psi$ masses, therefore larger couplings are needed in order to compensate for the reduction of the available phase space.

\begin{figure}[t!]
\begin{center}
\includegraphics[width=6cm]{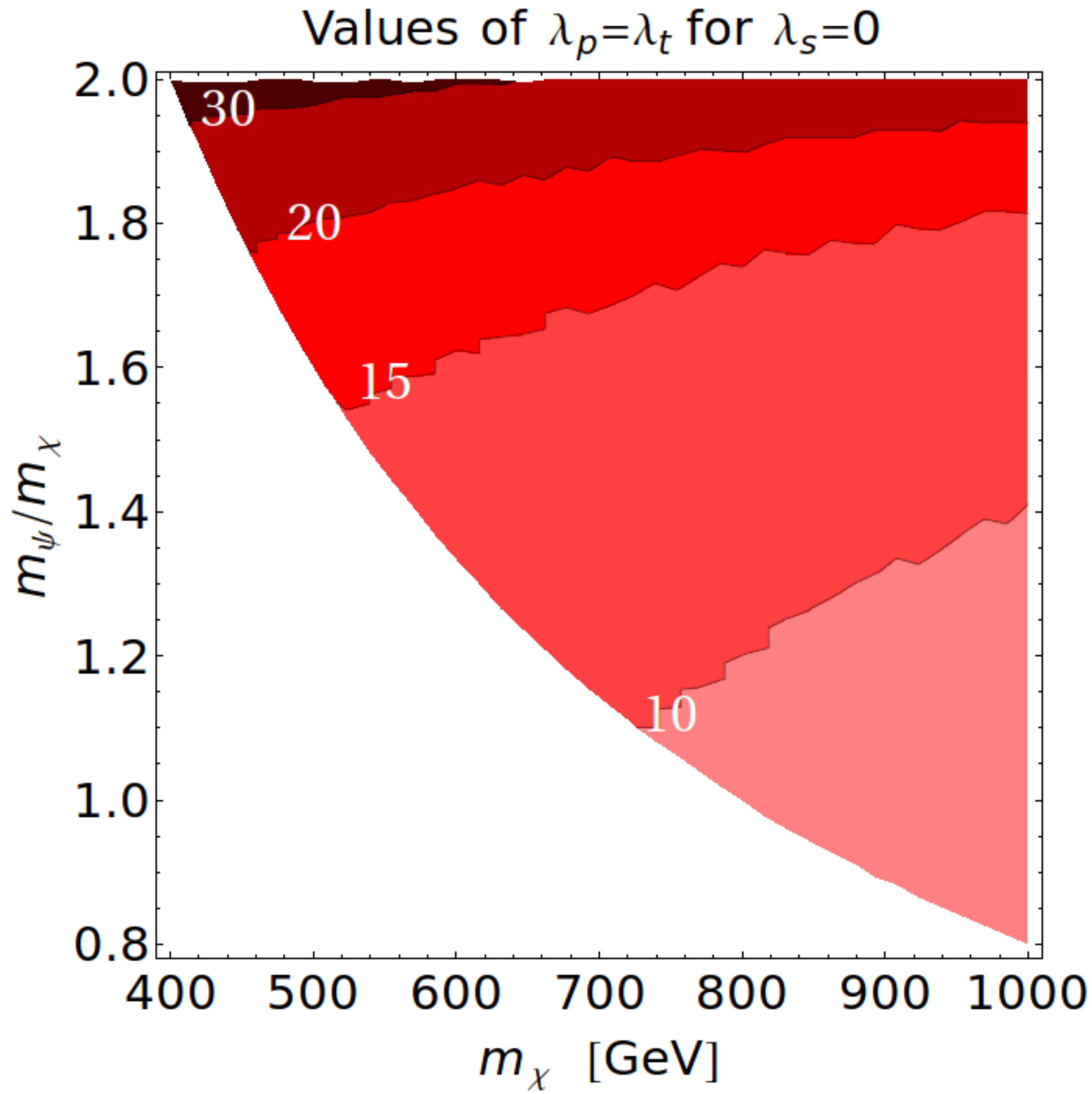}\qquad\includegraphics[width=6cm]{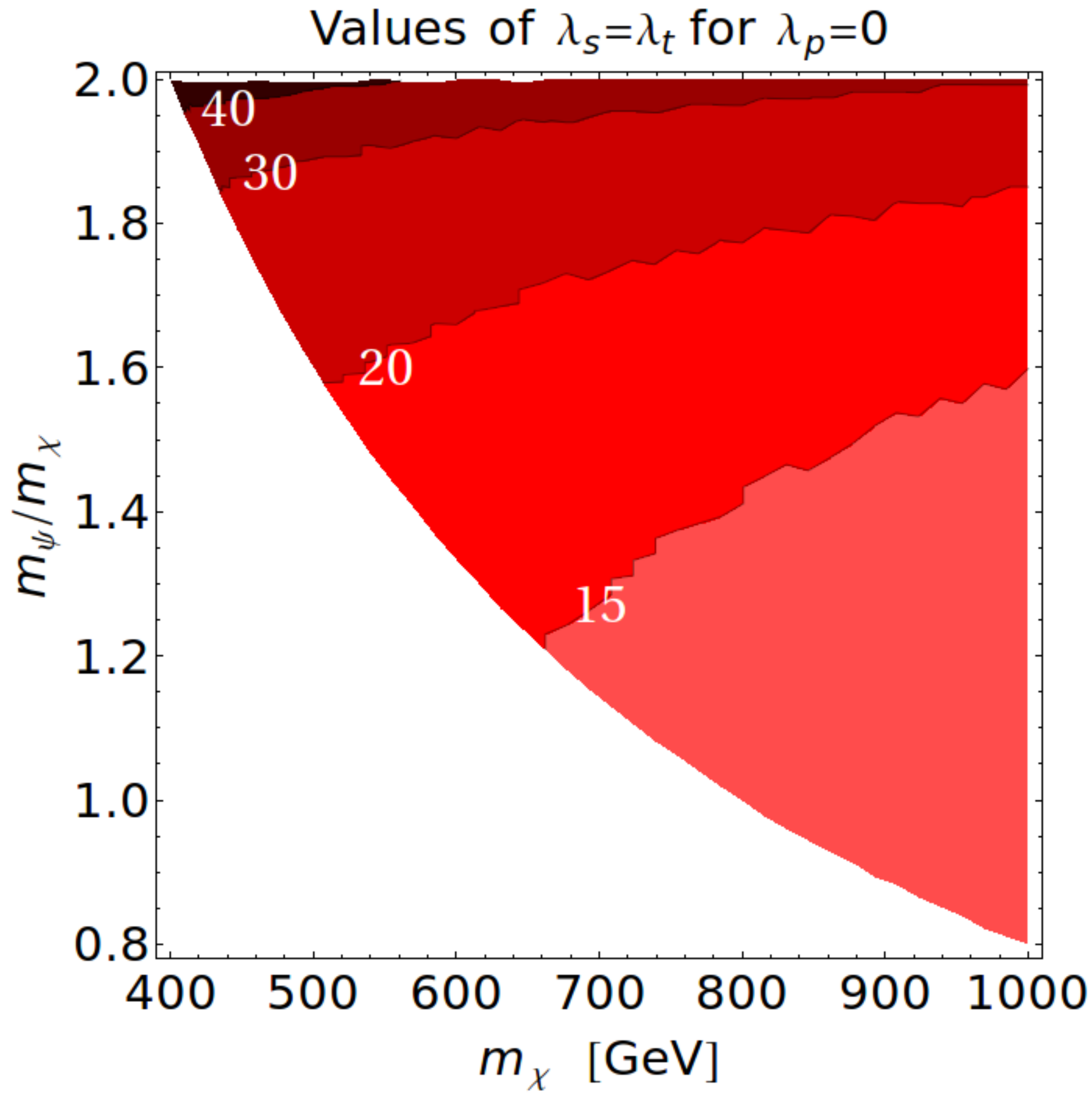}
\end{center}
\caption{\sl \textbf{\textit{Contour level for the couplings $\boldsymbol{\lambda_p=\lambda_t}$ (left) and $\boldsymbol{\lambda_s=\lambda_t}$ (right) needed for generating the DM relic density in the $\boldsymbol{[m_\psi/m_\chi,\,m_\chi]}$ plane.}}
The white lower left part corresponds to $m_\psi \leq 800$ GeV, which is excluded by the LHC.}
\label{Fig:GrapheRest1}
\end{figure}

Combining pseudoscalar and $t$-channel ($\lambda_p = \lambda_t $ and $\lambda_s=0$) or scalar and $t$-channel ($\lambda_s = \lambda_t $ and $\lambda_p=0$), we find the pattern that would be expected and inferred from the limiting cases analyzed above: 
the pseudoscalar dominates over the t-channel, and the t-channel dominates over the scalar one. This is illustrated in figure~\ref{Fig:GrapheRest1}.

The results in this subsection are in line with those obtained in other EFT approaches~\cite{Cao:2009uw,Zheng:2010js,Cheung:2012gi}.

\subsection{Baryon asymmetry}

Now we want to impose the constraint from the BAU, on top of the DM relic abundance already 
considered in the previous subsection. The final baryon asymmetry, that has to match the measured
 value, results from competing processes: CP-violating DM annihilations generate an asymmetry,
  while washout processes tend to deplete it.
  
\begin{figure}[t!]
\begin{center}
\includegraphics[width=10cm]{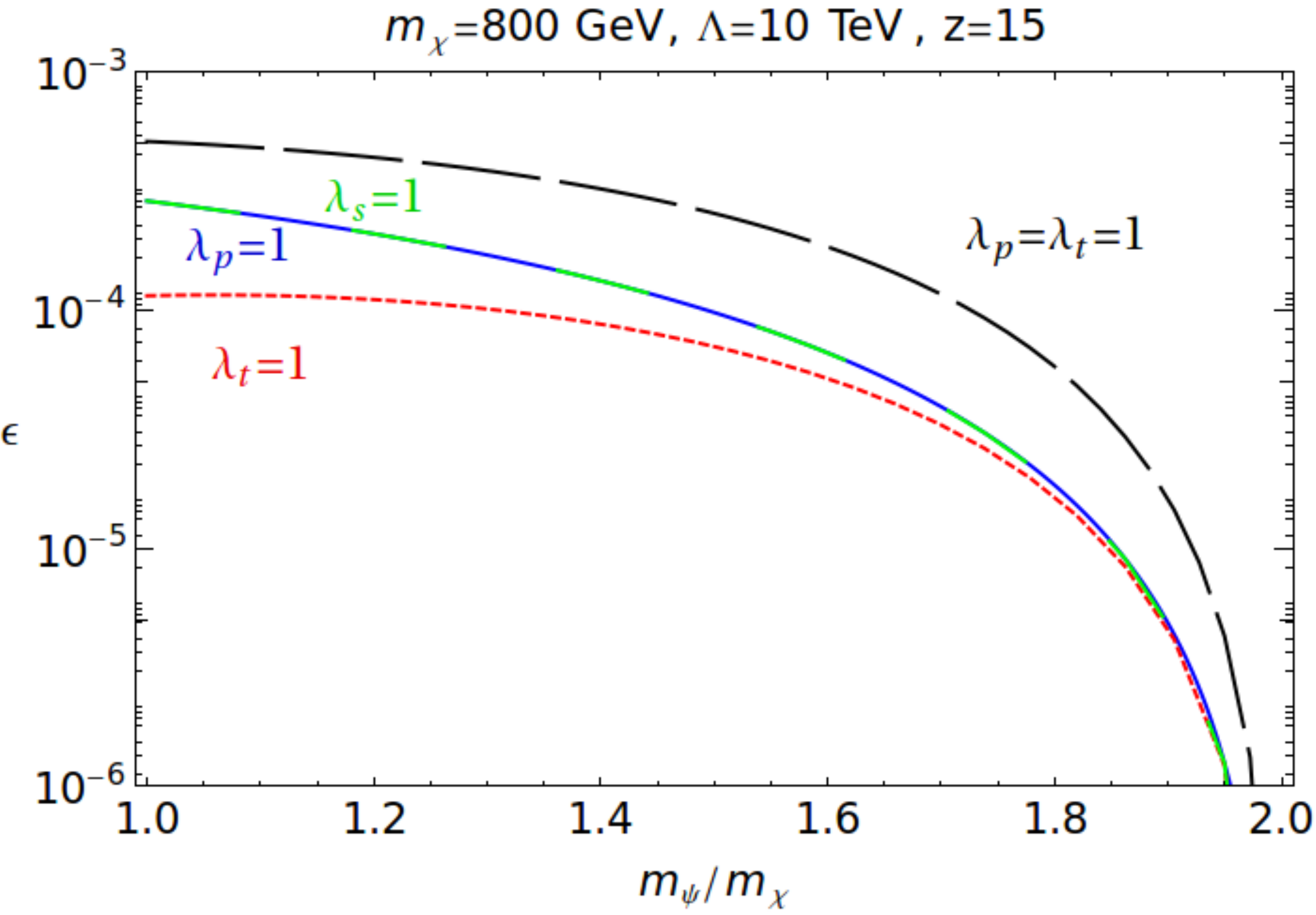}
\end{center}
\vspace{-0.8cm}
\caption{\sl \textbf{\textit{Comparing $\boldsymbol{\epsilon}$'s for various channels.}} The green dashed line is for $\lambda_s =1$ and $\lambda_{p}=\lambda_{t}=0$, which equals the case where $\lambda_p=1$ and $\lambda_{s}=\lambda_{t}=0$ depicted with the blue solid line.
 The dotted red line is for $\lambda_t =1$ and $\lambda_{s}=\lambda_{p}=0$, corresponding to a pure $t$-channel, but also to a $t$-channel plus scalar $s$-channel. The black long-dashed line stands for the case $\lambda_p=\lambda_t=1$ and $\lambda_s=0$.}
\label{fig:epsiloncompare}
\end{figure}

 Let us first consider the asymmetry, $\epsilon$, defined in eq.~\eqref{eq:epsilon} as the difference between the rate for 
DM annihilation into $\psi \ubar$ and the rate for the annihilation into $\psidag \ubardag$, normalized to the 
sum of the rates. As we explain in Appendix~\ref{app:epsilon}, after we define the couplings $\lambda_s
$, $\lambda_p$ and $\lambda_t$ by setting equalities among the a priori different $\lambda_i$'s, we 
need to make the choice of assigning the CP-violating phases to some of the couplings.
The most economical choice is to assign just one phase to the pure washout coupling, $\lambda_{\rm 
WO}$:
\begin{equation}
\lambda_\text{WO}=\vert \lambda_\text{WO} \vert\, e^{i\,\delta} \equiv \lambda_9 = \lambda_{11}\,.
\end{equation}
With this choice, we have (see eq.~\eqref{epsts})
\be \label{eq:epsprop}
 \epsilon\propto \vert \lambda_{\rm WO} \vert^2 \sin (2\,\delta)\,.
 \ee
To further simplify the analysis, we make the assumption that the phase is large and we 
set $\delta = \pi / 4$. This is not in conflict with current electric dipole moments (EDMs) measurements 
that would be the most constraining for these phases, given that the lowest order contribution to EDMs in 
these models is at three loops~\cite{Cui:2011ab} and so very suppressed\footnote{Although the 
diagrams in our EFT contributing to EDMs are different than the ones in~\cite{Cui:2011ab}, they are still 
three-loop suppressed.}. In figure~\ref{fig:epsiloncompare} we show $\epsilon$ for various limiting cases.
Note that the pseudoscalar and scalar channels give higher values of $\epsilon$ than the $t$-channel. In the cases with $\lambda_t=0$, the dependence on $\lambda_{s}$ and $\lambda_{p}$ cancels out in eq.~\eqref{epsts}, so that the scalar and the pseudoscalar $s$-channel asymmetries are equal.

\begin{figure}[!h]
\begin{center}
\includegraphics[width=10cm]{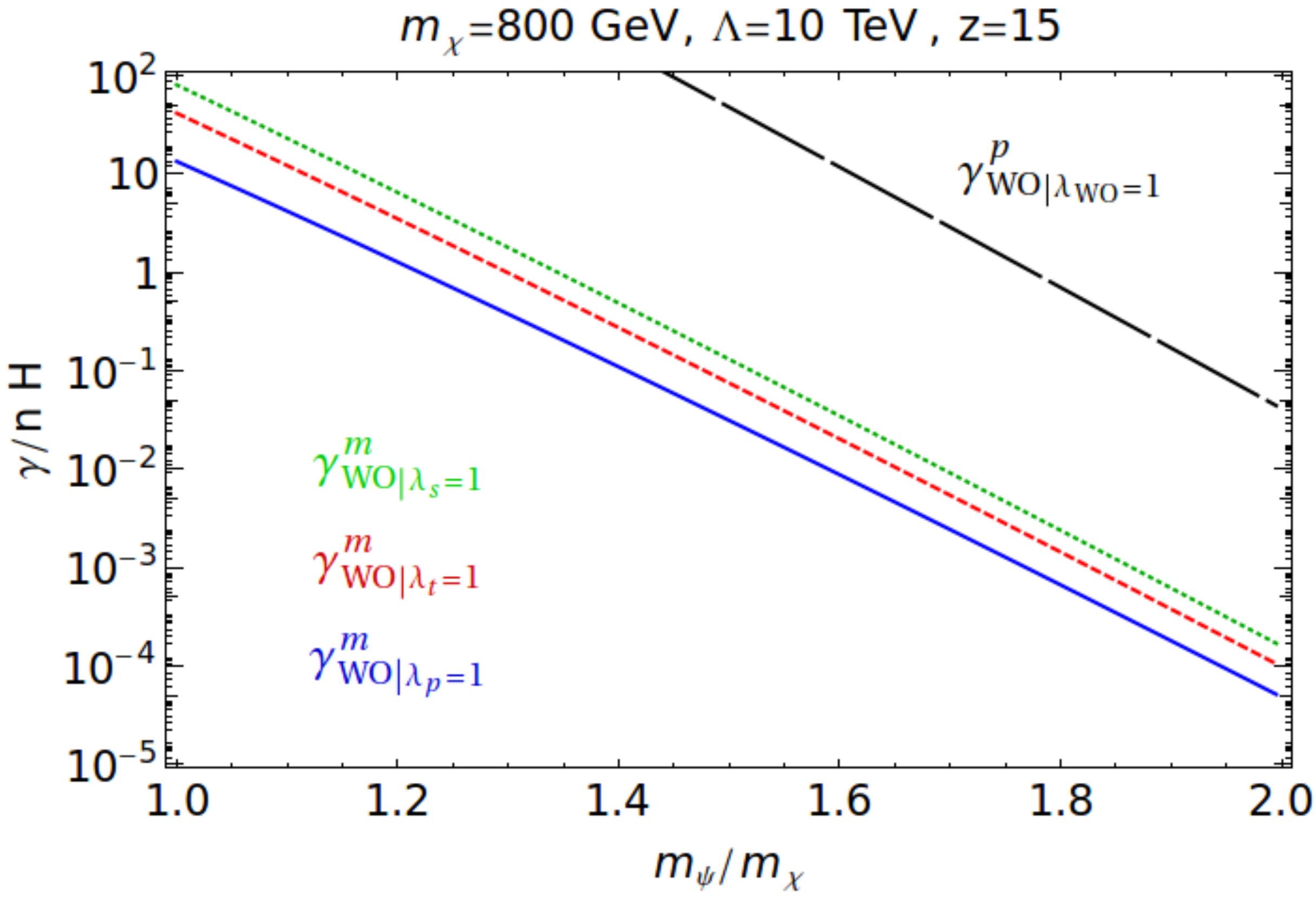}
\end{center}
\caption{\sl \textbf{\textit{Washout rates as a function of $\boldsymbol{m_\psi/m_\chi}$.}}
The rates are normalized to $n_\chi^{eq}(z)H(z)$. The color code is the same as in figure~\ref{Fig:Rates}: blue is for the pseudoscalar $s$-channel, red for the $t$-channel and green for the scalar $s$-channel. The black line here stands for the pure washout rate. For illustration, all respective couplings have been set to $1$.}
\label{washout}
\end{figure}

\begin{figure}[!h]
\begin{center}
\includegraphics[width=6cm]{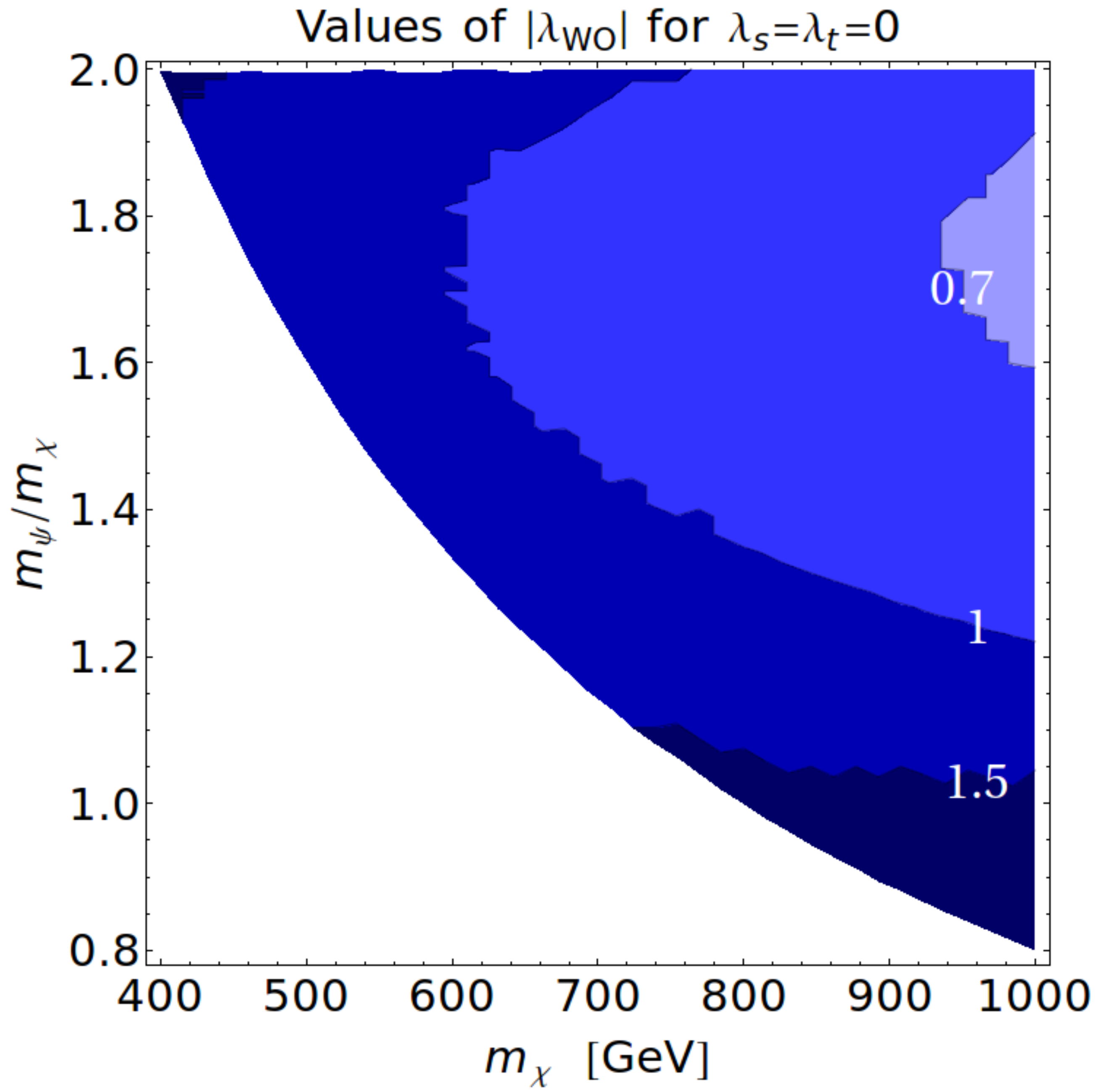}\qquad\includegraphics[width=6cm]{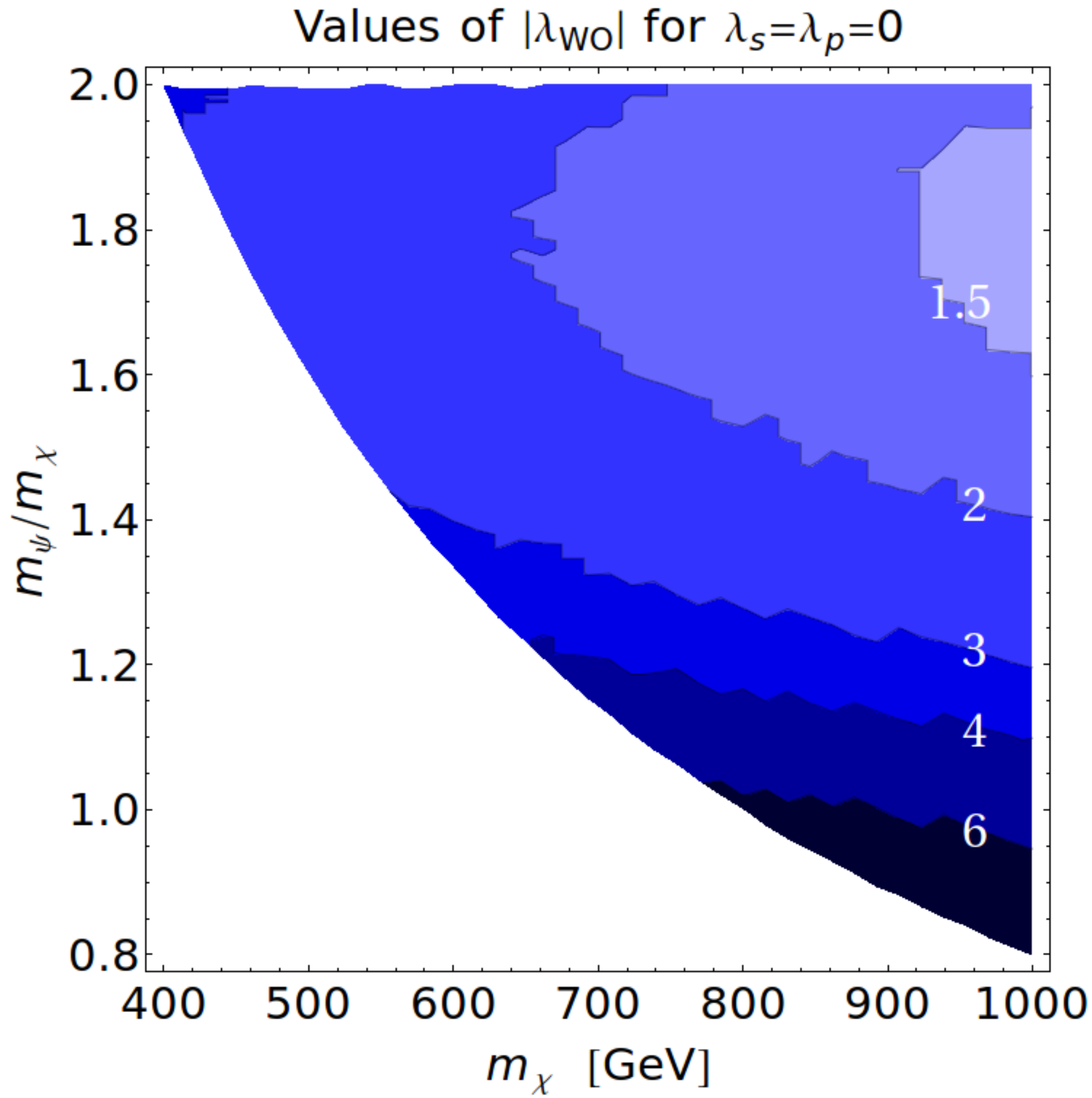}\\
\includegraphics[width=6cm]{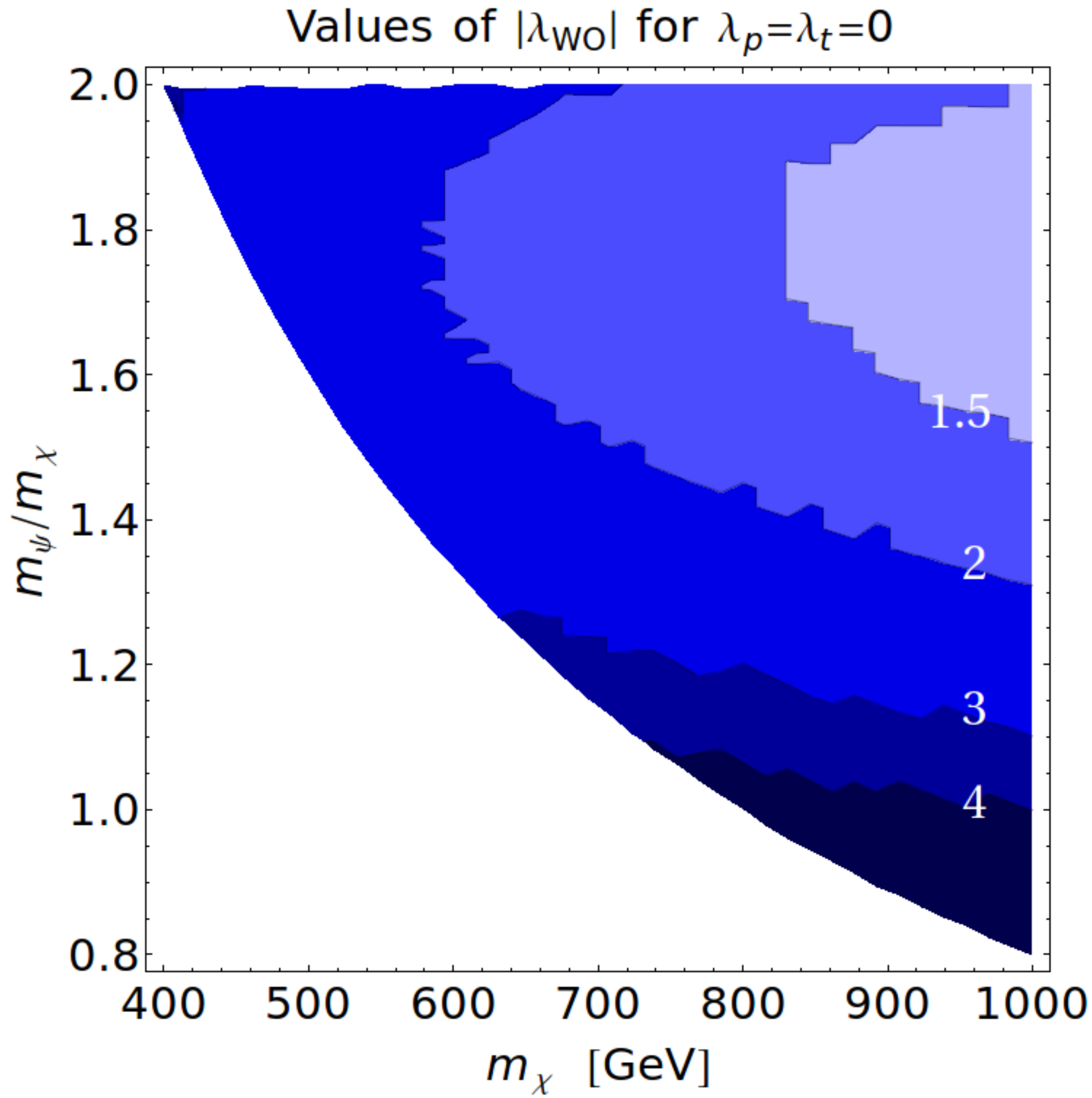}
\end{center}
\vspace{-0.8cm}
\caption{\sl \textbf{\textit{Contour levels for the modulus of the coupling $\boldsymbol{\lambda_\text{WO}}$ needed for generating the measured BAU, in the $\boldsymbol{[m_\psi/m_\chi,\,m_\chi]}$ plane.}}
We display the pseudoscalar (upper left pane), $t$-channel (upper right pane) and scalar (lower pane) cases.}
\label{Fig:GrapheLWO}
\end{figure}

  Next, let us consider the washout processes. The same couplings $\lambda_s$, $\lambda_p$ and 
$ \lambda_t$, that enter in the annihilations and have already been constrained by requiring the correct 
  DM relic density, also contribute to the mixed washout. In figure~\ref{washout} we show how the rates 
  for the mixed washout $\gamma_{\rm WO}^{m}$ compare in the three limiting cases (again normalized to $n_\chi^{eq}(z)H(z)$). Besides the very strong suppression of these 
  rates with increasing $m_\psi/m_\chi$ from kinematic closure, it is interesting to note that the pseudoscalar
  case (blue line) gives the least washout. Combined with the high annihilation rate and with the high values $\epsilon$, as shown in 
  figures~\ref{Fig:Rates} and \ref{fig:epsiloncompare}, this makes such channel the most promising for achieving WIMPy baryogenesis.
  The pure washout rate $\gamma_{\rm WO}^{p}$ is also depicted in figure~\ref{washout}, where we set all the couplings equal to 1 for illustration and comparison. It is important to keep in mind that each washout rate scales as the corresponding coupling $\lambda_i$ to the fourth power. We have already seen that the DM relic abundance requires the couplings $\lambda_s$, $\lambda_p$ and $\lambda_t$ to be at least of order 10. In figure~\ref{Fig:GrapheLWO} we see that $\lambda_{\rm WO}$, instead, is typically of order 1. As a consequence, the mixed washout rates are enhanced by a factor of $10^4$ compared to pure washout rates in our processes, thus they are dominant.

In figure~\ref{Fig:GrapheLWO} we depict contour levels for the modulus of the coupling $\lambda_\text{WO}$ needed for generating the measured value for the BAU, in the plane $[m_\psi/m_\chi,\,m_\chi]$, again for the limiting cases considered before.
Although not shown in the plots, the values for $\lambda_p$, $\lambda_s$ and $\lambda_t$ are fixed to reproduce the correct DM relic density.
It is worth noticing that, except for the white region already ruled out by the LHC, the rest of the parameter space could give rise to the correct DM relic density and the baryon asymmetry, for reasonable values for the couplings.
The behavior of $|\lambda_\text{WO}|$ can be understood as follows.
Recall that $\epsilon \propto |\lambda_{\rm WO}|^2 m_\chi^2$, while the mixed washout, which dominates over the pure one, is proportional to $\lambda_{p,\,s,\,t}^4$. For a fixed value of $m_\chi$, increasing $m_\psi$, both $\epsilon$ and the washout decrease. But while the washout rates decrease quickly, $\epsilon$ goes down slowly for $m_\psi / m_\chi \lesssim 1.8$, so $|\lambda_{\rm WO}|$ has to decrease in this direction in order to not overproduce the asymmetry. For $m_\psi / m_\chi \gtrsim 1.8$ the washout processes are not important anymore and $\epsilon$ would become too small, thus $|\lambda_{\rm WO}|$ has to invert the trend and start increasing. 
For a fixed $m_\psi / m_\chi$ ratio smaller values of $|\lambda_{\rm WO}|$ are needed when $m_\chi$ increases.

\begin{figure}[!h]
\begin{center}
\includegraphics[width=6cm]{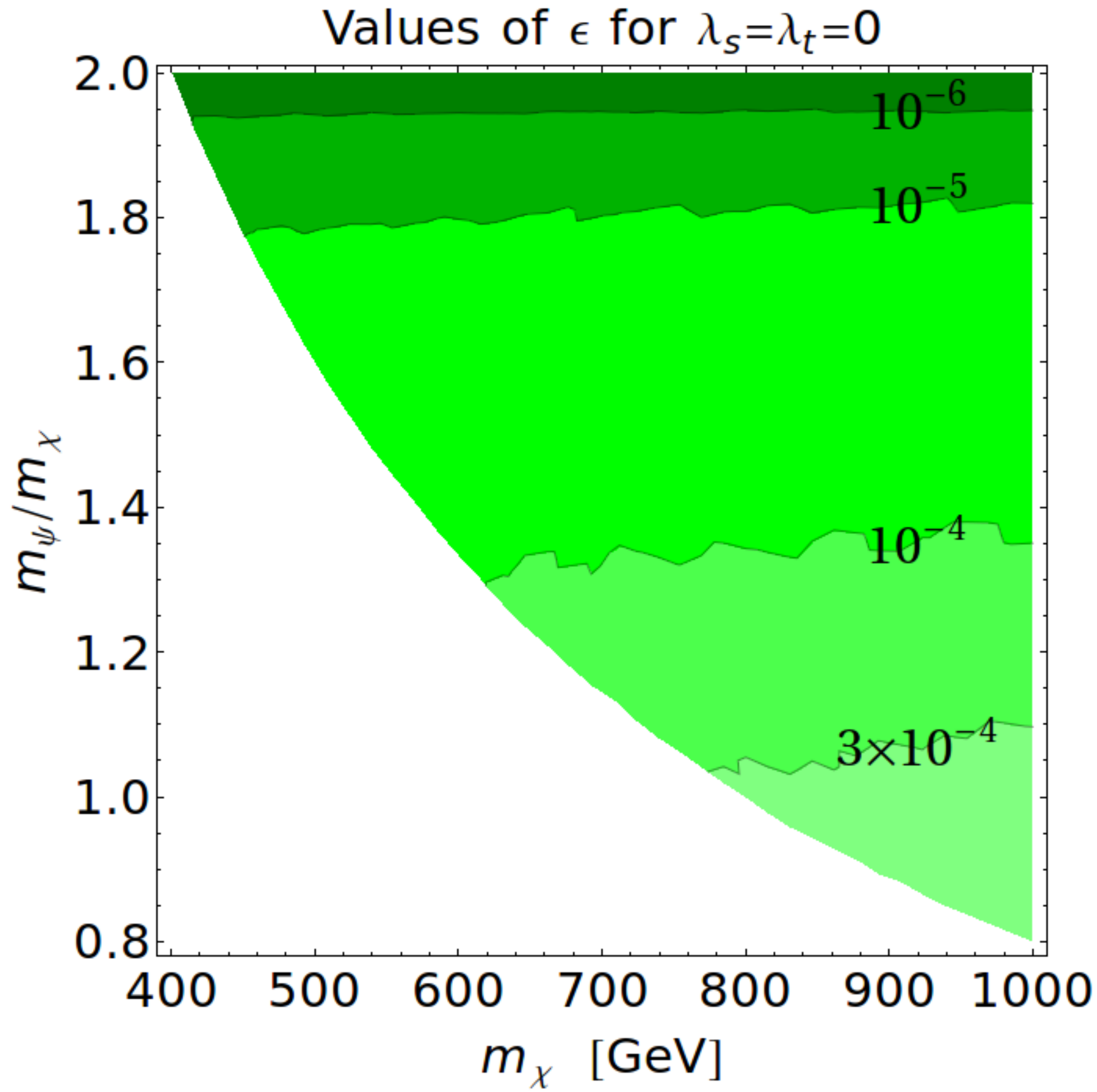}\qquad\includegraphics[width=6cm]{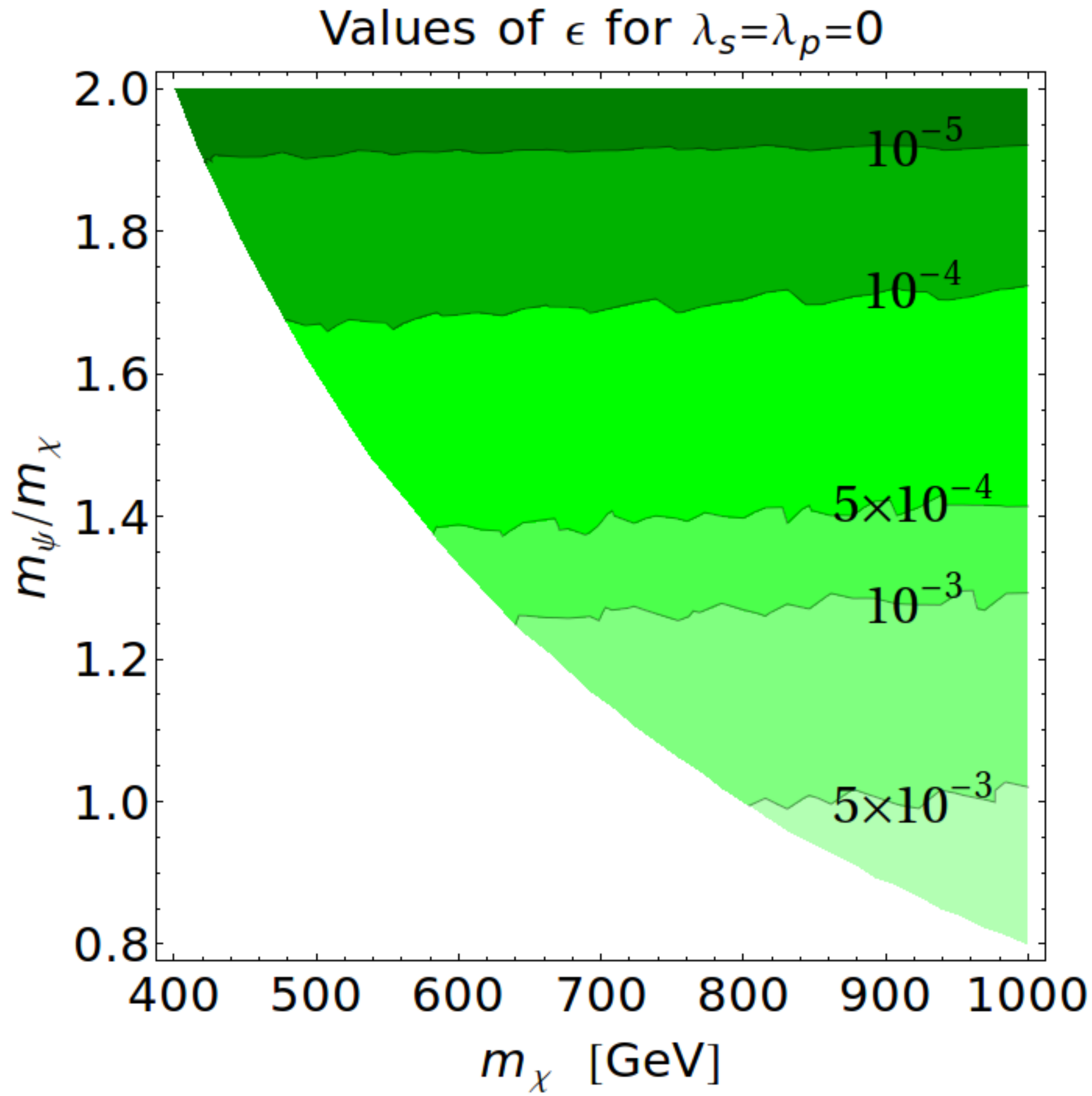}\\
\includegraphics[width=6cm]{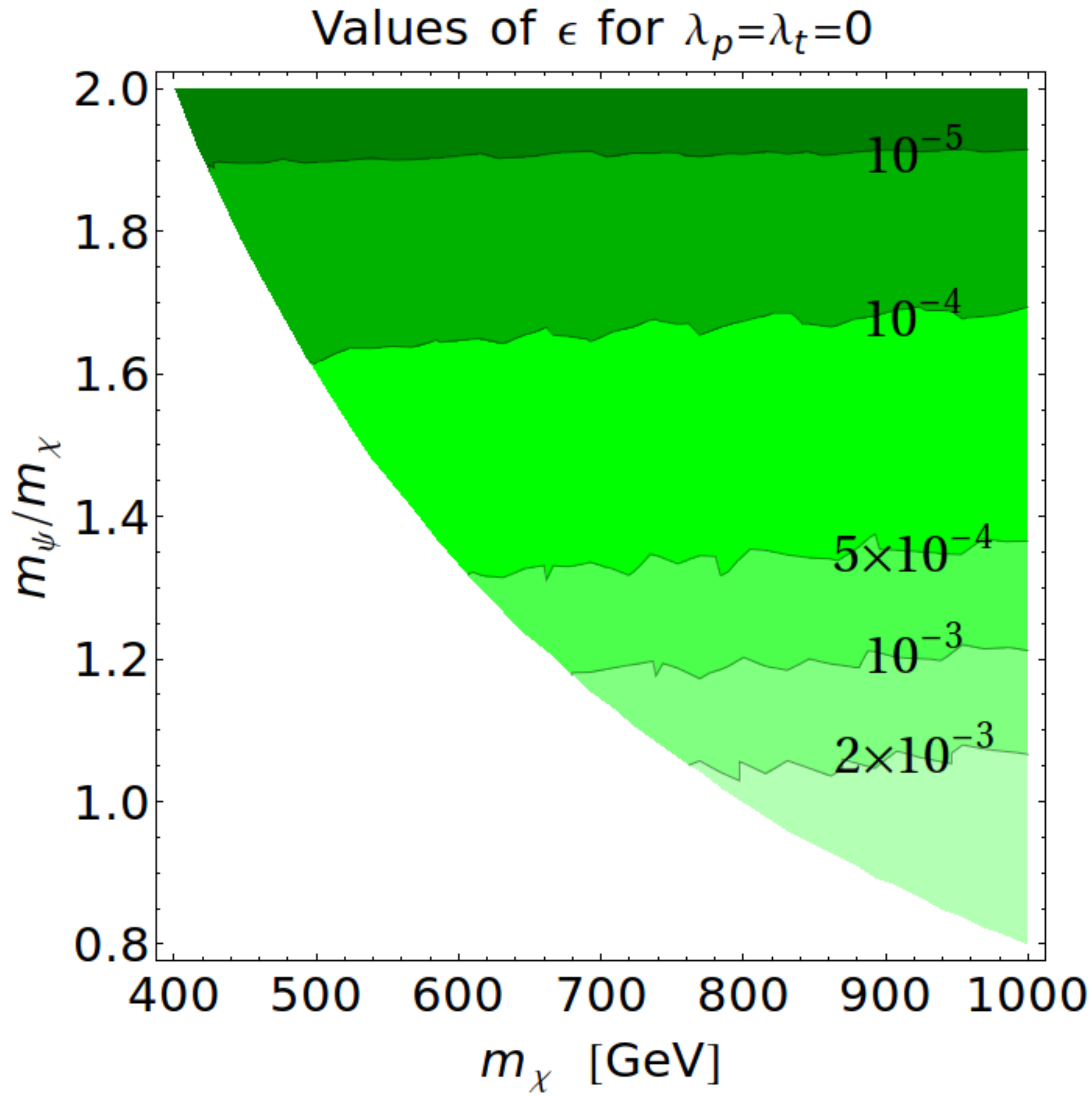}
\end{center}
\vspace{-0.8cm}
\caption{\sl \textbf{\textit{Contour levels for $\boldsymbol{\epsilon}$.}}
We display the pseudoscalar (upper left pane), $t$-channel (upper right pane) and scalar (lower pane) cases.}
\label{Fig:eps}
\end{figure}

Figure~\ref{Fig:eps} shows contour levels for $\epsilon$ (in the low temperature limit) generated when imposing both the DM relic density and BAU constraints.
The parameter $\epsilon$ roughly follows the same behavior as if we were keeping $\lambda_\text{WO}$ constant: it decreases with the increase of $m_\psi/m_\chi$ (see equation~\eqref{epsts}).

\begin{figure}[h!]
\begin{center}
\includegraphics[width=6cm]{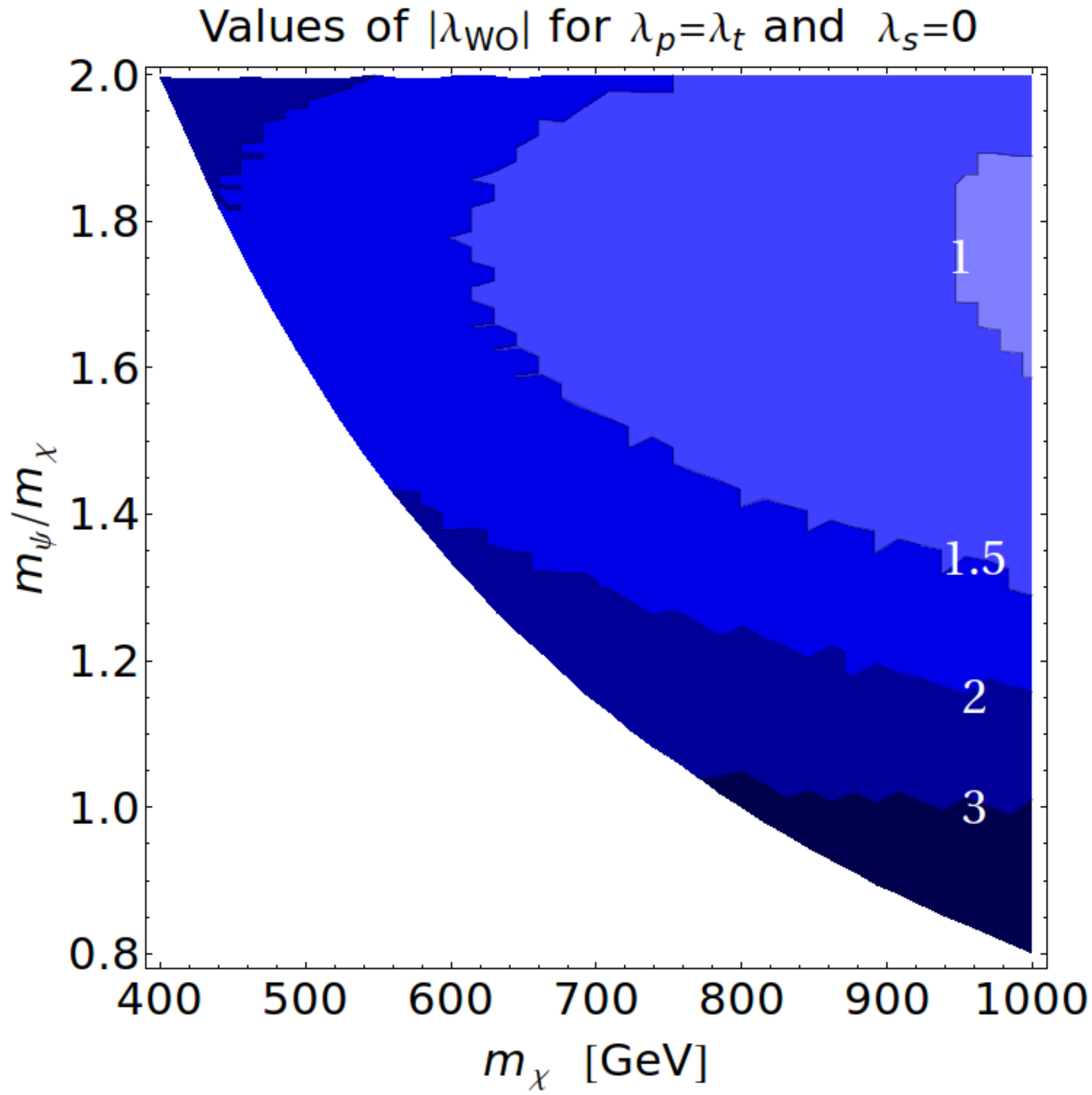}\qquad\includegraphics[width=6cm]{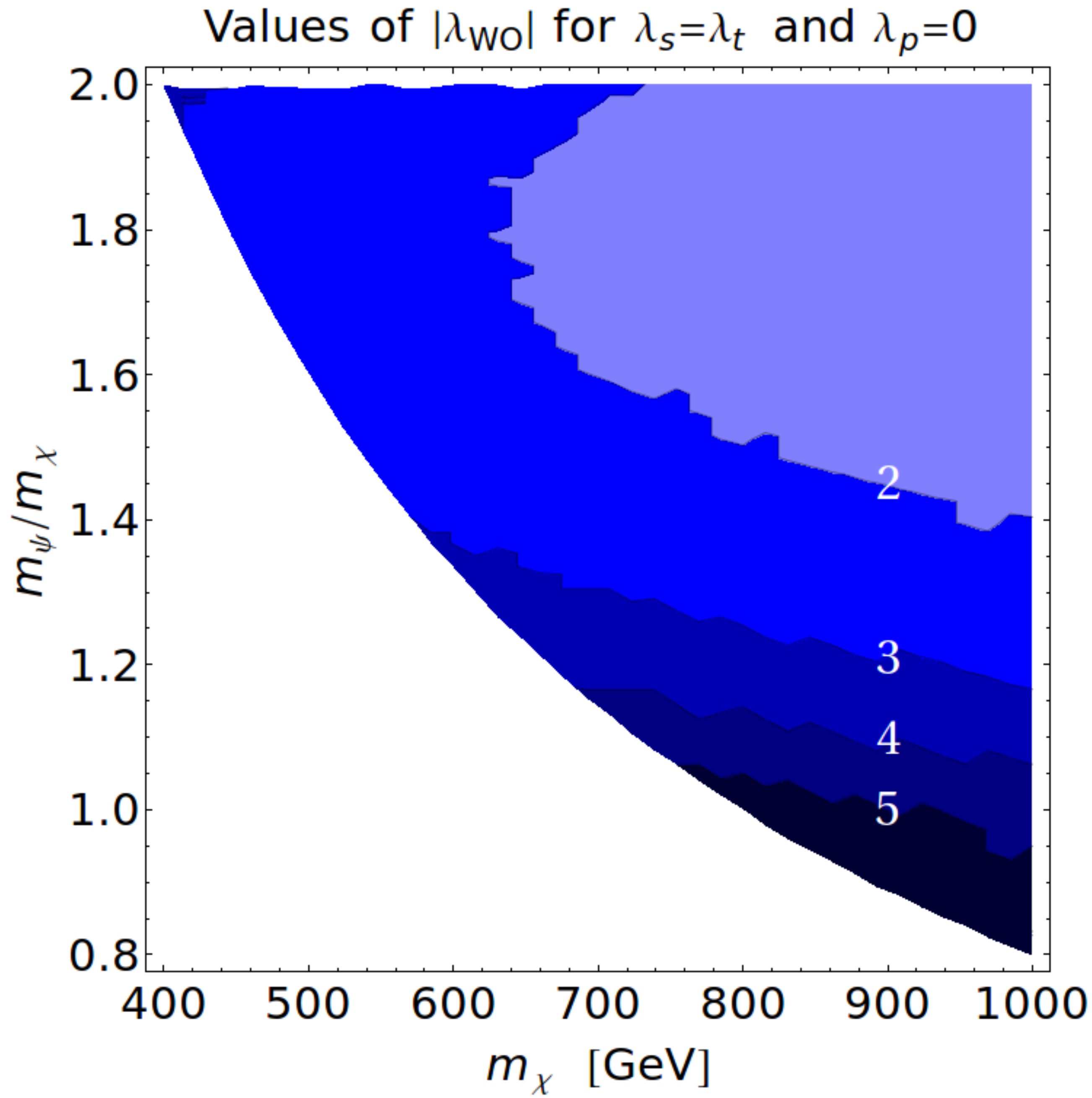}
\includegraphics[width=6cm]{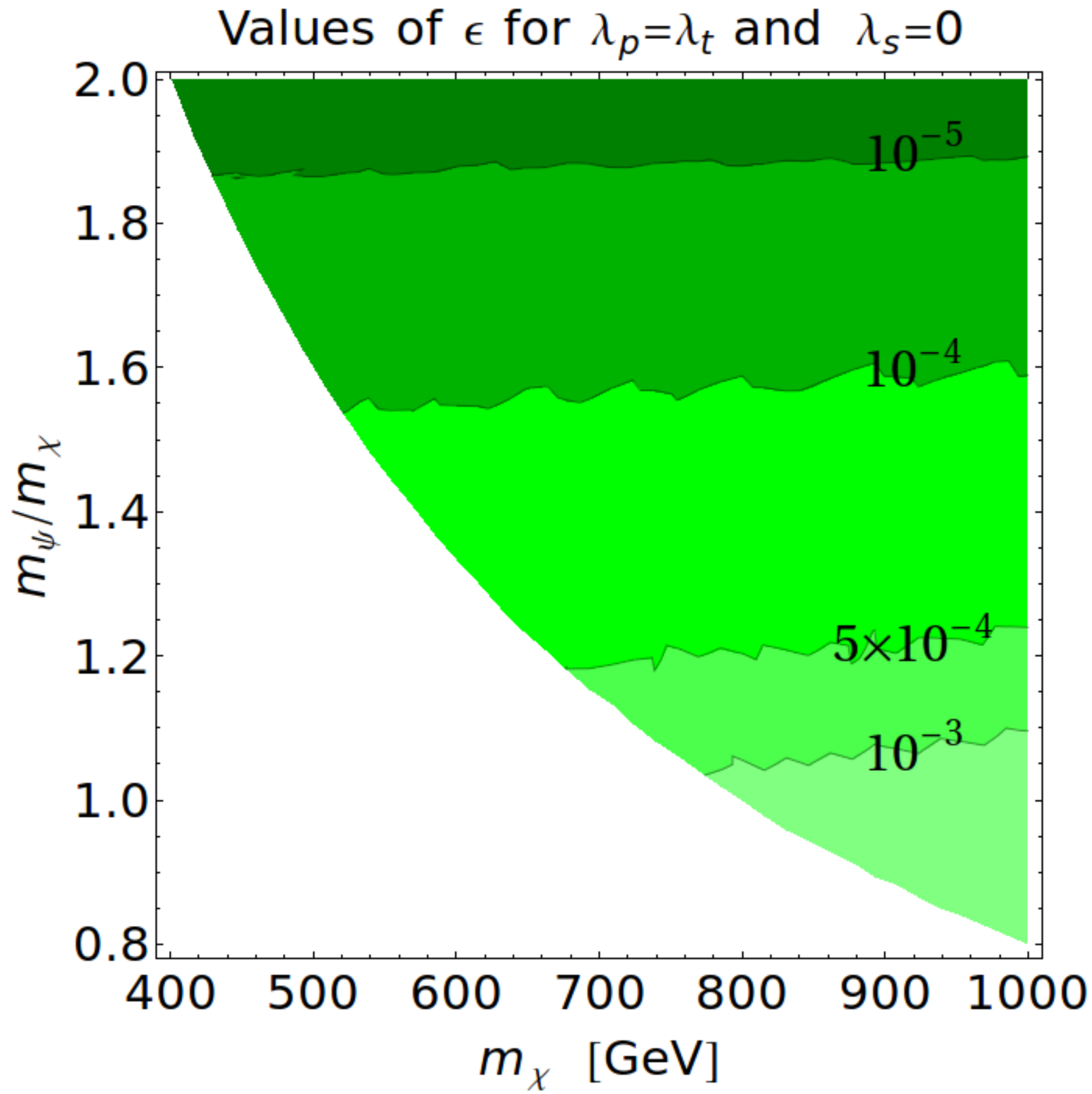}\qquad\includegraphics[width=6cm]{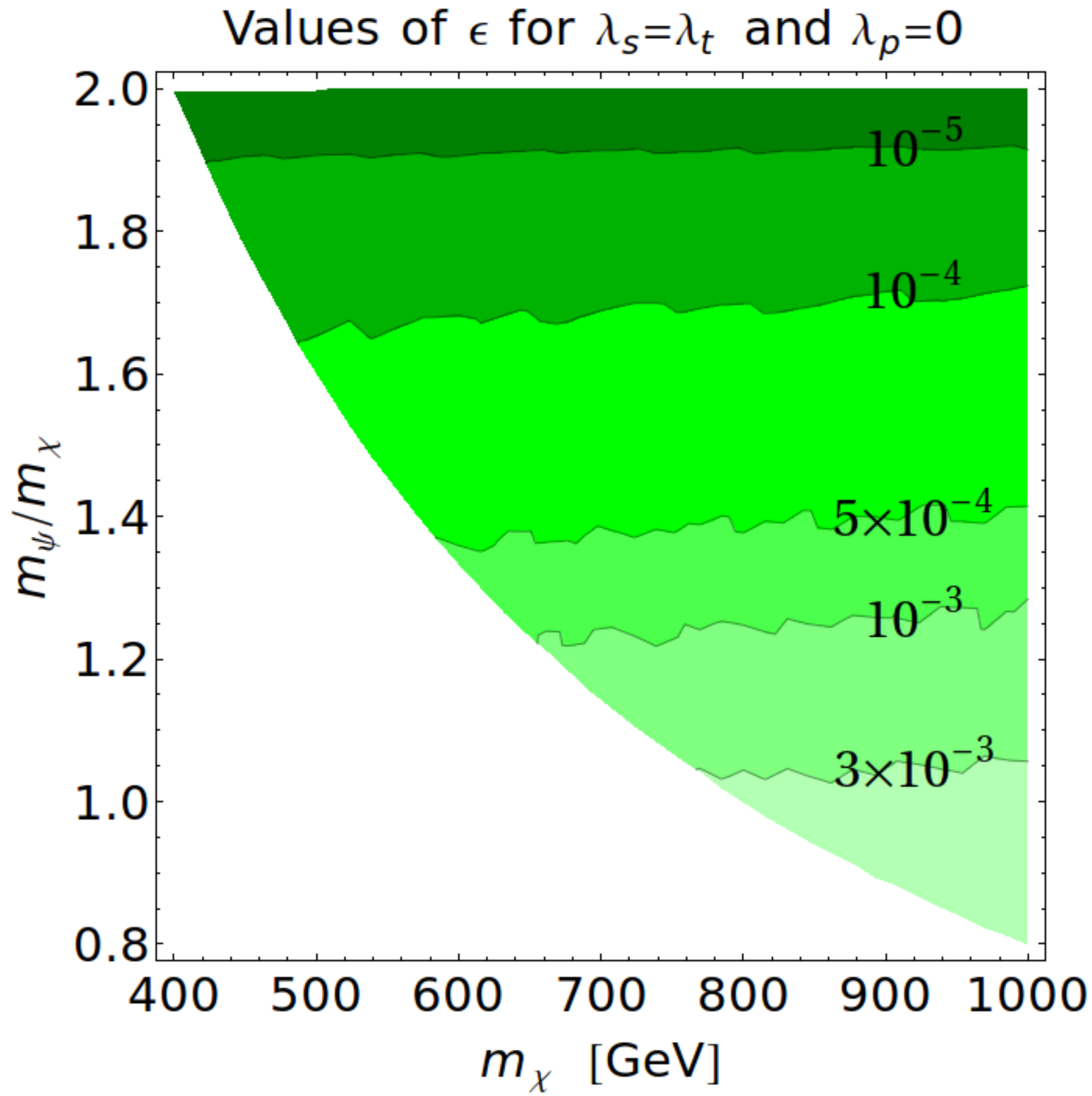}
\end{center}
\vspace{-0.8cm}
\caption{\sl \textbf{\textit{Contour levels for the modulus of the coupling $\boldsymbol{\lambda_\text{WO}}$ (top panels) and the parameter $\epsilon$ (bottom panels) needed for generating both the DM relic abundance and the BAU.}}
Left panels show $\lambda_p=\lambda_t$ with $\lambda_s=0$, while on the right we set $\lambda_s=\lambda_t$ with $\lambda_p=0$.}
\label{Fig:GrapheRest2}
\end{figure}

For completeness, we display the remaining two limiting cases, where $\lambda_p=\lambda_t$ with $\lambda_s=0$, and $\lambda_s=\lambda_t$ with $\lambda_p=0$ in figure~\ref{Fig:GrapheRest2}.
The constraints on $|\lambda_\text{WO}|$ and on $\epsilon$ are shown in the upper and lower panels respectively.
Whereas we had a clear dominance of the pseudoscalar channel in the determination of the DM abundance, we see in the figures of this subsection that the parameters $\epsilon$ and $\lambda_{\rm WO}$ do not vary significantly among the three limiting cases.

\section{Constraints from dark matter direct detection}\label{sec:direct}

We have two operators in our effective Lagrangian that contribute to the direct detection of DM at tree level:
\be \label{eq:7&8}
\frac{1}{\Lambda^2}\left[\lambda_7^2 (X\ubar)(\Xdag \ubardag) + \lambda_8^2 (\Xbar \ubar)(\Xbardag \ubardag)\right] +{\rm h.c.}\,.
\ee
When eq.~\eqref{eq:7&8} is translated into four-component-spinor language 
\begin{eqnarray}
&&\frac{\lambda_8^2-\lambda_7^2}{4 \Lambda^2} \left[(\bar\chi \gamma^\mu \chi) (\bar U \gamma_\mu U) + (\bar\chi \gamma^\mu \chi) (\bar U \gamma_\mu \gamma_5 U)\right] \nonumber\\
&&\qquad\qquad + \frac{\lambda_8^2+\lambda_7^2}{4 \Lambda^2} \left[(\bar\chi \gamma^\mu \gamma_5 \chi) (\bar U \gamma_\mu U) + (\bar\chi \gamma^\mu \gamma_5 \chi) (\bar U \gamma_\mu \gamma_5 U)\right]\,,
\end{eqnarray}
we see clearly that the first term (vector coupling) gives a spin-independent (SI) contribution, the last term (axial-vector coupling) a spin-dependent (SD) contribution, while the second and third terms are velocity suppressed and can be neglected. Note that the SI term is proportional to the difference of the couplings, while the SD one to the sum. 

The operators in eq.~\eqref{eq:7&8} are also responsible for DM annihilation into a quark plus an anti-quark. This annihilation channel, which does not contribute to the asymmetry, would be competing with the one into quark plus exotic anti-quark. We want the former to be suppressed with respect to the latter, in order to generate the correct BAU. Therefore, even strict bounds on the couplings $\lambda_7$ and $\lambda_8$ from direct detection, would not challenge these models. Put another way, for the WIMPy baryogenesis to work, $\lambda_7$ and $\lambda_8$ must be suppressed with respect to $\lambda_s$, $\lambda_p$ and $\lambda_t$, which would explain why no signal has been seen so far in direct detection experiments.

\begin{figure}[!h]
\begin{center}
\includegraphics[width=12cm]{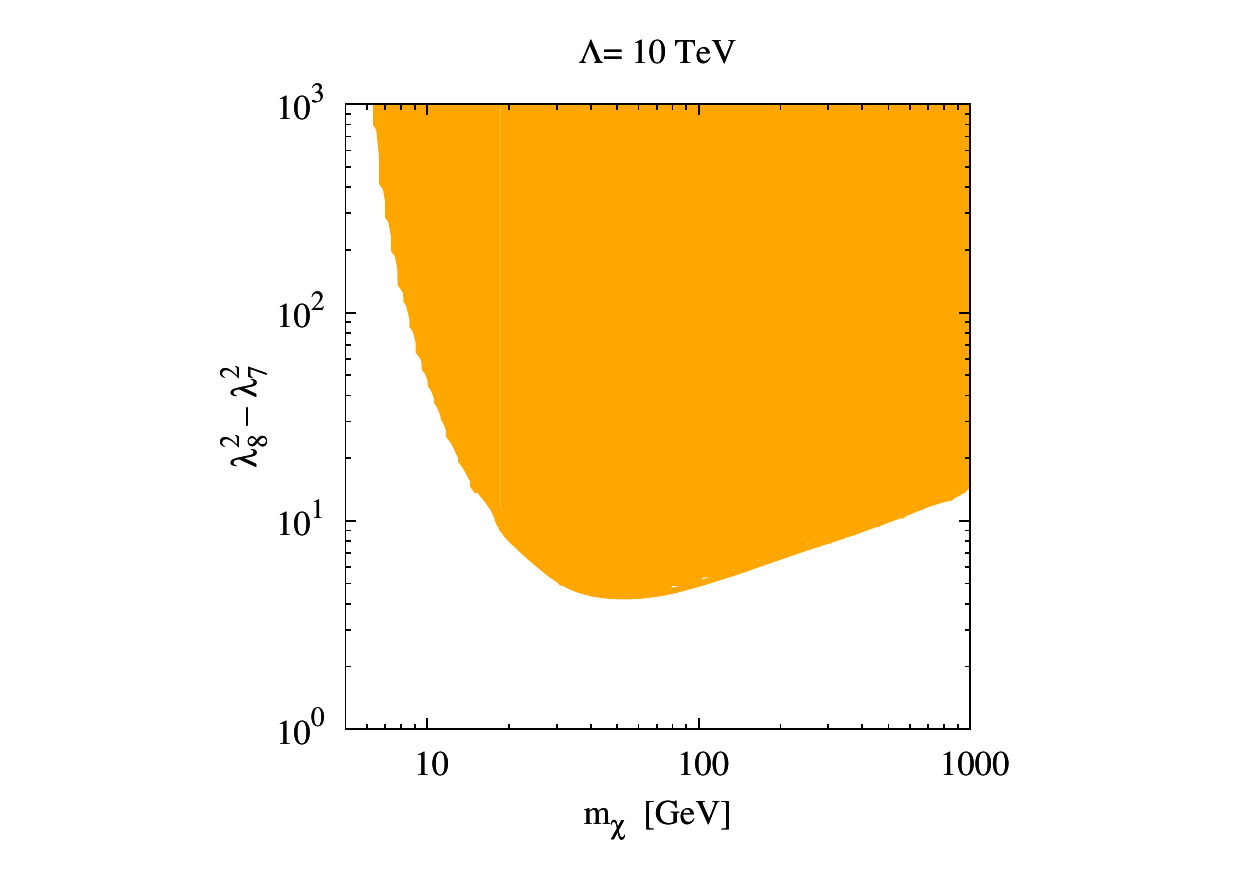}
\end{center}
\vspace{-0.8cm}
\caption{\sl \textbf{\textit{XENON100 sensitivity.}} The upper orange region on
the parameter space $[\lambda_8^2-\lambda_7^2,\,m_\chi]$ is
already ruled out by the latest XENON100~\cite{Aprile:2012nq} measurements.}
\label{Fig:xenon}
\end{figure}

Let us take a look at the bounds, starting with the case that $\lambda_7$ and $\lambda_8$ are different and consider the SI contribution. The WIMP-nucleon cross section is given by
\begin{equation}
\sigma_0=\frac{\mu_N^2\,B_N^2}{\pi}\,.
\label{xenon}
\end{equation}
In equation~\eqref{xenon}, $B_N\equiv\alpha_u\,(A+Z)+\alpha_d\,(2\,A-Z)$, $A$ is the atomic mass number, $Z$ is the atomic number and $\mu_N$ is the WIMP-nucleus reduced mass.
$\alpha_u$ and $\alpha_d$ correspond to the couplings between DM and the up- and down-quark respectively: $\alpha_u=\frac{\lambda_8^2-\lambda_7^2}{4\,\Lambda^2}$, $\alpha_d=0$. 
Figure~\ref{Fig:xenon} depicts the impact of the XENON100 exclusion limits, after $225$ live days of data~\cite{Aprile:2012nq}, on the plane $[\lambda_8^2-\lambda_7^2,\,m_\chi]$.
The upper orange region is ruled out.
For the relevant mass region of the present study (i.e. $m_\chi\gtrsim 400$ GeV) the combination of the couplings $\lambda_7^2 - \lambda_8^2$ has to be smaller than $\sim 10$. 

If $\lambda_7 = \lambda_8$, there is no SI contribution, but the SD one is at its maximum. The SD limits are a few orders of magnitude weaker\footnote{This is certainly true for heavy DM, which is our case, where the best limits are from direct detection experiments, see e.g.~\cite{Akimov:2011tj, Felizardo:2011uw}. For light DM one would get better SD limits from colliders~\cite{Bai:2012xg, Goodman:2010ku, Fox:2011pm}.}
than SI limits, and are of no interest for the present study, given that they are weaker than the bound from the validity of the EFT approach, $\lambda_i < 100$.

Constraints on the couplings $\lambda_s$, $\lambda_p$ and $\lambda_t$ would be more interesting, because they play a crucial role in the generation of the baryon asymmetry and the determination of the DM relic abundance. The lowest order contribution to the direct detection involving these couplings is naïvely expected to be at one loop. 

\begin{figure}[h!]
\begin{center}
\includegraphics[width=0.7\textwidth]{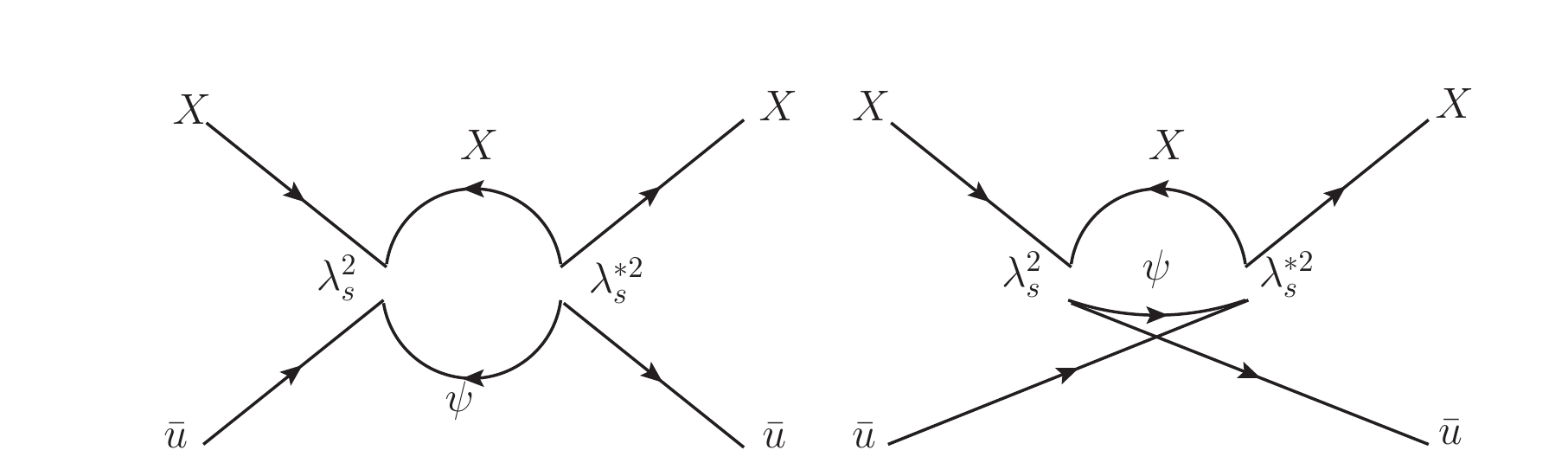}
\end{center}
\vspace{-0.8cm}
\caption{\sl \textbf{\textit{Diagrams for direct detection at one loop.}} In the limit of zero external momenta the sum of these two diagrams vanishes.
}
\label{Fig:oneloops}
\end{figure}

Let us take a closer look at this statement and let us consider first the situation where only the $s$-channel is turned on ($\lambda_t=0$). There are two one-loop diagrams, as shown in Figure~\ref{Fig:oneloops}. We show the $\lambda_s$ coupling in the figure, but if we replaced it with $\lambda_p$ the following argument would still apply. 
It is easy to see that, in the limit of zero external momenta, appropriate for direct detection, the sum of the two diagrams vanishes. This is understood just by looking at the fermionic propagator for $\psi$ in the loop, the direction of which in the second diagram is opposite to the one in the first, thus giving a relative minus sign. This is enough to conclude that the contribution to the cross section is not only loop suppressed but also velocity suppressed.
As a consequence current direct detection experiments place virtually no limits on the coupling $\lambda_s$. The exact same conclusion holds for the pseudoscalar coupling $\lambda_p$.

For the $t$-channel ($\lambda_t \neq 0$, $\lambda_s = \lambda_p = 0$), there is no analogous obvious cancellation at the level of one-loop Feynman diagrams, but it turns out that there is only a SD contribution, which is loop suppressed. Since, as we said, the SD limits are not even important for the tree-level couplings, we are definitely far from putting constraints on $\lambda_t$ with direct detection.



\section{Summary and discussion}\label{sec:summary}
In this work we have investigated whether a general class of WIMPy baryogenesis models is viable, after experimental constraints are taken into account. 
Our models are based on the same mechanism and on the same external particles as in Ref.~\cite{Cui:2011ab}.
However, by following an EFT approach and writing down a complete list of four-fermion operators, we extend and generalize their study, considering all the possible DM annihilation channels.
The models considered here require the presence of a heavy fermion, $\psi$, which is crucial to the success of the whole mechanism. Because $\psi$ is colored, the LHC represents an excellent laboratory for testing these models. Although we have not yet studied in detail possible collider signals, current LHC searches already put a lower bound of 800 GeV on the mass of $\psi$, which in turn directly translates into a lower bound of 400 GeV on the DM mass. With the impressive pace at which the LHC and the ATLAS and CMS collaborations are operating, this bound can increase relative quickly, pointing to even higher masses, or, in a better (luckier) scenario, a heavy colored fermion could be discovered soon, which would provide a hint that these models could be realized in nature indeed.

In this work we focused mainly on the cosmological aspects and we examined in some detail the constraints from the measured DM relic density and BAU. We considered three different channels for DM annihilation into a quark and an exotic antiquark: scalar and pseudoscalar $s$-channel, and $t$-channel. We found the pseudoscalar channel to be the most promising: it has the highest annihilation cross section, the lowest washout cross section and it generates a large asymmetry $\epsilon$. This combination results in lower values of the coupling $\lambda_p$, compared to $\lambda_s$ and $\lambda_t$, and in the most efficient production of the BAU. In the spirit that lower rather than higher values of the couplings are generally preferred in the EFT, our analysis, in all cases, points toward a high DM mass, between 800 GeV and 1 TeV, and  a small hierarchy between $\psi$ and $\chi$, $m_\psi \lesssim 1.4\, m_\chi$  (see the figures in Section~\ref{sec:cosmology}).

We also considered bounds from direct detection. These constrain only two operators that would be responsible for the annihilation of DM into a pair of quarks. Given that this channel does not contribute to the generation of the baryon asymmetry, we want it to be suppressed anyway. In this sense such bounds do not challenge these models at all. There are in principle one-loop diagrams, involving the couplings that also enter the generation of the asymmetry, that could contribute to the direct detection cross section. We showed that they are not only loop suppressed, but also velocity suppressed. Thus, this scenario is out of reach for current direct detection experiments.

Since CP violation is a crucial ingredient in these models, one has to worry that the physical phases are not too constrained. As already pointed out in~\cite{Cui:2011ab}, it seems that the strongest constraints on the phases would come from EDMs measurements. For the models of~\cite{Cui:2011ab}, the lowest order contribution to the neutron EDM only appears at three loops, and as a consequence their phases are not much constrained at all. In our EFT context, the diagrams contributing to the neutron EDM are slightly different, but it is still true that the lowest order contribution appears at three loops, so we reach the same conclusion: the CP violating phases are not much constrained by current experiment, so we have the freedom of taking them quite large.

There are more aspects of WIMPy baryogenesis that we have not touched here, but we believe are worth investigating, such as indirect detection signals, and the possibility that the light stable state $n$, into which the exotic quark $\psi$ decays, could constitute the extra degree of radiation at BBN time. There is also more work related to model-building that should be done. For example one could think about UV completions of the models presented here, and it would be interesting to explore other discrete symmetry groups that would work with the mechanism.

\acknowledgments
We would like to thank Arindam Chatterjee, Manuel Drees, Herbi Dreiner, Marta Losada, Juan Racker, Witold Skiba and Sean Tulin for helpful discussions.
NB and LU are supported by the DFG TRR33 ``The Dark Universe''.
The work of FXJM is funded by  the Fundaç\~{a}o para a Ci\^encia e a Tecnologia (FCT, Portugal) through the projects  CERN/FP/123580/2011 and CFTP-FCT Unit 777, which are partially funded through POCTI (FEDER).

\appendix

\section{The complete list of dimension six operators} \label{app:list}
Given the particle content in Table~\ref{tab:content}, we want to write down all the possible dimension six operators. Our Lagrangian has the form
\be \label{eq:effLapp}
\mathcal{L} \supset \frac{1}{\Lambda^2} \sum_i \lambda_i^2 \mathcal{O}_i.
\ee
In this paper we set the scale $\Lambda$ to 10 TeV. The reader should keep in mind that varying the couplings $\lambda_i$'s is equivalent to varying the scale $\Lambda$, the only measurable quantity in the EFT being the combination $\lambda_i / \Lambda$.

The couplings $\lambda_i$'s are complex in principle. As we show in Appendix~\ref{app:epsilon}, we definitely need some of them to be complex in order to have the CP violation necessary for the generation of a baryon asymmetry.

The strategy for writing down all possible operators is:
\begin{itemize}
\item write down all the operators consistent with the symmetries,
\item use Fierz identities to reduce to a linearly independent basis.
\end{itemize}

\paragraph{Operators linear in $\psi$ and $\ubar$}
\be
\begin{split}
\lambda_1^2 (XX)(\psi \ubar) + \lambda_2^2 (\Xbar \Xbar)(\psi \ubar) + \lambda_3^2 (\Xdag \Xdag)(\psi \ubar) +\lambda_4^2 (\Xbardag \Xbardag)(\psi \ubar) \\ 
+ \lambda_5^2 (\Xbardag \psibardag)(X \ubar) + \lambda_6^2 (\Xdag \psibardag)(\Xbar \ubar) \\
+ {\rm h.c.}
\end{split}
\ee
These operators are relevant for the annihilation of DM, proceeding through an $s$-channel (first line) or through a $t$-channel (second line).

\paragraph{Operators bilinear in $\ubar$}
\be
\lambda_7^2 (X\ubar)(\Xdag \ubardag) + \lambda_8^2 (\Xbar \ubar)(\Xbardag \ubardag) +{\rm h.c.}
\ee
These contribute to the DM annihilation into a pair of quarks and to the tree-level direct detection cross section.

\paragraph{Operators bilinear in $\psi$ and $\ubar$}
\be
\lambda_9^2 (\psi \psi)(\ubar\ubar)+\lambda_{10}^2 (\psi \ubar)(\psidag \ubardag) + \lambda_{11}^2 (\psibardag \psibardag)(\ubar \ubar) + \lambda_{12}^2 (\psibardag \ubardag)(\psibar \ubar) + {\rm h.c.}
\ee
The operators 9 and 11 contribute to the washout cross section.

\paragraph{Operators bilinear in $\psi$}
\be \label{eq:psipsi}
\begin{split}
\lambda_{13}^2 (X\Xbar)(\psi \psibar) + \lambda_{14}^2 (\Xdag \Xbardag)(\psi\psibar)  \\
+ \lambda_{15}^2(X\psi)(\Xbar\psibar)  + \lambda_{16}^2(\Xdag \psibardag)(X\psibar) + \lambda_{17}^2 (\Xbardag\psibardag)(\Xbar\psibar) + \lambda_{18}^2 (\Xdag\psidag)(X\psi) +\lambda_{19}^2 (\Xbardag\psidag)(\Xbar\psi) \\
+ {\rm h.c.}
\end{split}
\ee
These contribute to the DM annihilation only when $m_\psi<m_\chi$. If one wants to successfully generate a baryon asymmetry through the annihilation of DM, these operators need to be suppressed with the respect to the ones linear in $\psi$ and $\ubar$.

\paragraph{Operator for the decay of $\psi$}
\be
\lambda_{20}^2 (\psibar \bar d)(\bar d n) + {\rm h.c.}
\ee
Depending on the charge assignment of $n$, see Table~\ref{tab:content}, this operator either conserves baryon number or violates it.

\section{The role of the discrete symmetry} \label{app:discrete}
Discrete symmetries, such as $R$-parity in SUSY or KK-parity in Universal Extra Dimensions, just to mention two popular examples, are a generic feature of models with DM candidates. In this appendix we show that, considering the Abelian discrete group $\mathbb{Z}_n$, we need at least $n=4$ for our models.

Let us start with a generic $\mathbb{Z}_n$ and let us assign the following charges
\be
Q_X = \exp \left(\frac{2\pi i}{n} q_X\right), \qquad Q_\psi = \exp \left(\frac{2\pi i}{n} q_\psi\right), \qquad Q_{\ubar} = \exp \left(\frac{2\pi i}{n} q_{\ubar}\right).
\ee
To avoid proton decay we charge all the SM quarks, but keep all the leptons neutral. This way a baryon cannot decay into mesons and/or leptons.
Then we have the following requirements:
\begin{itemize}
\item dark matter, $X$, has to be stable. This implies that $q_X \neq 0$;
\item we want to avoid decays of the exotic quark, $\psi$, into SM quarks only,
in order not to spoil the asymmetry. Given that we have charged the SM quarks, we can avoid such dangerous decays by keeping $\psi$ neutral, $q_\psi = 0$;
\item to generate the asymmetry, we need DM to annihilate both into $\psi +\ubar$ and their conjugates. This requires two operators, $(XX)(\psi \ubar)$ and $(XX)(\psidag \ubardag)$.   
\end{itemize}
The last requirement imposes the following conditions on the charges
\beqn
2q_X + q_{\ubar} &=& 0 \, ({\rm mod} \ n), \\
2q_X - q_{\ubar} &=& 0 \, ({\rm mod} \ n),
\eeqn
from which we find either $q_X=0$ or $q_X = n/4$. As we said, for DM stability, $X$ has to be charged, so the first solution is not acceptable. We conclude that
\be \label{eq:charges}
q_X = n/4, \qquad q_{\ubar} = n/2.
\ee
Given that the $q_i$'s must be integers, this implies $n=4k$, with $k$ an integer. This proves that the discrete group has to be $\mathbb{Z}_{4k}$, the smallest one being $\mathbb{Z}_4$.

The solution~\eqref{eq:charges} has an important implication: the discrete charge of $X$ is complex. This means that a Majorana DM does not work here. Therefore, the simple considerations that we have outlined force the {\em dark matter fermion to be Dirac} in our models.
Note that we do not expect this to be a generic feature of WIMPy baryogenesis models: it is conceivable that other non-Abelian discrete symmetries can stabilize DM without requiring it being Dirac.

The $\mathbb{Z}_4$ symmetry also prevents any coannihilation process between $\psi$ and $\chi$.

\section{Boltzmann equations}\label{app:BEs} 

The evolution of the DM, $\psi$ and baryon asymmetry number densities in the expanding Universe is governed by a set of Boltzmann equations.
Introducing the rescaled inverse temperature $z\equiv m_\chi / T$ and the comoving number densities $Y_{\xi}\equiv n_{\xi}(z) / s(z)$, $s(z)$ being the entropy density, we can write:
\begin{small}
\begin{eqnarray}
z\,s(z)\,H(z)\frac{d Y_\text{DM}}{dz} & =& - 2\,\left(\gamma_{\rm ann}^{\rm CPV}(z)+\gamma_{\rm ann}^{\rm CPC}(z)\right) \left( \left(\frac{Y_\text{DM}}{Y_\text{DM}^\text{eq}}\right)^2-1\right)\label{eqDM}\,,\\
z\,s(z)\,H(z)\frac{d Y_{\Delta u}}{dz} & =& \epsilon(z)\,\gamma_{\rm ann}^{\rm CPV}(z)\left(\left(\frac{Y_\text{DM}}{Y_\text{DM}^\text{eq}}\right)^2-1\right)\nonumber\\
&&-\left( \frac{Y_{\Delta u}}{Y_{u}^\text{eq}}-\frac{Y_{\Delta \psi}}{Y_{\psi}^\text{eq}}\right)\left(\frac{Y_\text{DM}}{Y_\text{DM}^\text{eq}} \gamma^{m}_\text{WO}(z)+2\,\gamma^{p}_\text{WO}(z)\right)\label{eqBAU}\,,\\
z\,s(z)\,H(z)\frac{d Y_{\Delta \psi}}{dz} & =& -\epsilon(z)\,\gamma_{\rm ann}^{\rm CPV}(z)\left(\left(\frac{Y_\text{DM}}{Y_\text{DM}^\text{eq}}\right)^2-1\right)+\left( \frac{Y_{\Delta u}}{Y_{u}^\text{eq}}-\frac{Y_{\Delta \psi}}{Y_{\psi}^\text{eq}}\right)\left(\frac{Y_\text{DM}}{Y_\text{DM}^\text{eq}} \gamma^{m}_\text{WO}(z)+2\,\gamma^{p}_\text{WO}(z)\right)\, \nonumber\\
&&-\gamma_{D}\left(\frac{Y_{\Delta \psi}}{Y_{\psi}^\text{eq}}+2 \frac{Y_{\Delta d}}{Y_{d}^\text{eq}}\right)\label{eqdp}\,,\\
z\,s(z)\,H(z)\frac{d Y_{\Delta d}}{dz} & =&-2\gamma_{D}\left(\frac{Y_{\Delta \psi}}{Y_{\psi}^\text{eq}}+2 \frac{Y_{\Delta d}}{Y_{d}^\text{eq}}\right)\label{eqdd} \,,
\end{eqnarray}
\end{small}
where $H(z)$ is the Hubble expansion rate.
The thermally averaged interaction rates are defined  for a $i\,j \leftrightarrow k\,l$ scattering by
\be \label{eq:rates}
\gamma^{i\,j}_{k\,l}(T)=\frac{T}{64\,\pi^4}\int_{s_\text{inf}}^{\infty} ds\,\sqrt{s}\,K_{1}\left(\frac{\sqrt{s}}{T}\right)\,\hat{\sigma}^{i\,j}_{k\,l}(s)\,,
\ee
were the boundary is $s_\text{inf}= {\rm max}\left\lbrace (m_i+m_j)^2 ,\,(m_k+m_l)^2\right\rbrace$, and $\hat{\sigma}$ is the reduced cross sections are given in terms of the Mandelstam variable by
\be
\hat{\sigma}^{i\,j}_{k\,l}(s)=\frac{1}{8\,\pi\,s} \int_{t_{0}}^{t_{1}}dt \left\vert \mathcal{M}(i\,j\rightarrow k,\,l) \right\vert^2\,,
\ee
and the integration limits are given for example in the PDG~\cite{PhysRevD.86.010001}.

The Boltzmann equations are quadratic in the DM density $Y_\text{DM}=Y_{X}+Y_{\overline{X}}$, but we only expand them to first order in the asymmetries $\epsilon$ and $Y_{\Delta \alpha}=Y_\alpha-Y_{\overline{\alpha}}$, with $\alpha=u, d,\psi$, as these are expected to be small. Up to this approximation and given the list of operators above, these equations are the most general ones one can write.
They reflect the fact that while DM annihilates through many channels, only a part of them ($X X \to \psi \,\overline{u}$~+~h.c.) is CP-violating (CPV) and contributes to the generation of SM and exotic baryons asymmetries.
The CPV interactions are proportional to the $u$-quark number violating operators $\mathcal{O}_{1\dots 6}$, and we provide an approximate formula for $\gamma_{\rm ann}^{\rm CPV}$ in eq.~(\ref{eq:DMrate}). The CP-conserving annihilations involve the operators $\mathcal{O}_{7,\,8}$ and $\mathcal{O}_{13\dots 19}$. As the SM baryon-number is conserved by the latter processes, they do not contribute to the generation of the BAU.
In that respect, they constitute a competitive effect that reduces the WIMPy baryogenesis efficiency, so they need to be suppressed. We henceforth neglect them.
Further, the number asymmetries undergo  washout processes. The mixed washout, $\gamma^{m}_\text{WO}$ (operators $\mathcal{O}_{1\dots 6}$),  mix DM with $u$-$\psi$, whereas the pure washout, $\gamma^{p}_\text{WO}$ (operators $\mathcal{O}_{9}$ and $\mathcal{O}_{11}$), involves only $u$ and $\psi$.  We plot in figure~\ref{Fig:washout} the rates for various limiting cases.

In equations~\ref{eqdp} and~\ref{eqdd}, we introduced the decays of $\psi$ into two down-quarks, parametrized by the rate $\gamma_D$.
This decay provides the only source of $d$-quark number violation.
Let us note that the decays of $\psi$ inject high-energy down-quarks in the thermal bath.
To avoid spoiling Big-Bang nucleosynthesis predictions (see, e.g.~\cite{Kawasaki:2004qu}), these decays have to be fast enough, with $\tau(\psi \to \bar{d}\,\bar{d}\,n) \lesssim 1$ s. This places constraints on $\lambda_n$ that are rather weak. In order to simplify the Boltzmann equations we can ask that $\psi$ be in thermal equilibrium up to the freeze-out of washout. Estimating $\psi$ decay rates by 
\begin{equation}
 \Gamma(\psi \to \bar{d}\,\bar{d}\,n) \simeq \frac{\lambda_n^4\,m_\psi^5}{2^{13}\,\pi^3\,\Lambda^4}\,,
\end{equation}
 we then obtain
\begin{equation}
 \Gamma(\psi \to \bar{d}\,\bar{d}\,n) \gtrsim H(z_{\rm WO}) \Rightarrow \lambda_n \gtrsim 8\cdot 10^{-3}\times \left(\frac{m_\chi}{1 \ \text{TeV}}\right)^{1/2}\left(\frac{m_\psi}{1.5 \ \text{TeV}}\right)^{-5/4}\,\left(\frac{\Lambda}{10 \ \text{TeV}}\right)\,.
\end{equation}
The effect of these fast decays are drastic for the $\psi$ abundance: every $\psi$ produced in $X$ annihilation immediately decays into two anti-down quarks, enforcing $\Delta n_\psi \simeq 0$, while the total baryon asymmetry yielded is approximately $\Delta n_u+\Delta n_d \simeq 3\,\Delta n_u$. 
At the end, the Boltzmann equations simplify to:
\begin{eqnarray}
z\,s(z)\,H(z)\frac{d Y_\text{DM}}{dz} & =& - 2\,\gamma_{\rm ann}^{\rm CPV}(z) \left( \left(\frac{Y_\text{DM}}{Y_\text{DM}^\text{eq}}\right)^2-1\right)\,, \\
z\,s(z)\,H(z)\frac{d Y_{\Delta u}}{dz} & =& \epsilon(z)\,\gamma_{\rm ann}^{\rm CPV}(z)\left(\left(\frac{Y_\text{DM}}{Y_\text{DM}^\text{eq}}\right)^2-1\right)\nonumber\\
&&- \frac{Y_{\Delta u}}{Y_{u}^\text{eq}}\left(\frac{Y_\text{DM}}{Y_\text{DM}^\text{eq}} \gamma^{m}_\text{WO}(z)+2\,\gamma^{p}_\text{WO}(z)\right)\,.
\end{eqnarray}
We numerically solve this set of equations.

\section{Calculation of the CP asymmetry} \label{app:epsilon}

\begin{figure}[t!]
\begin{center}
\includegraphics[width=\textwidth]{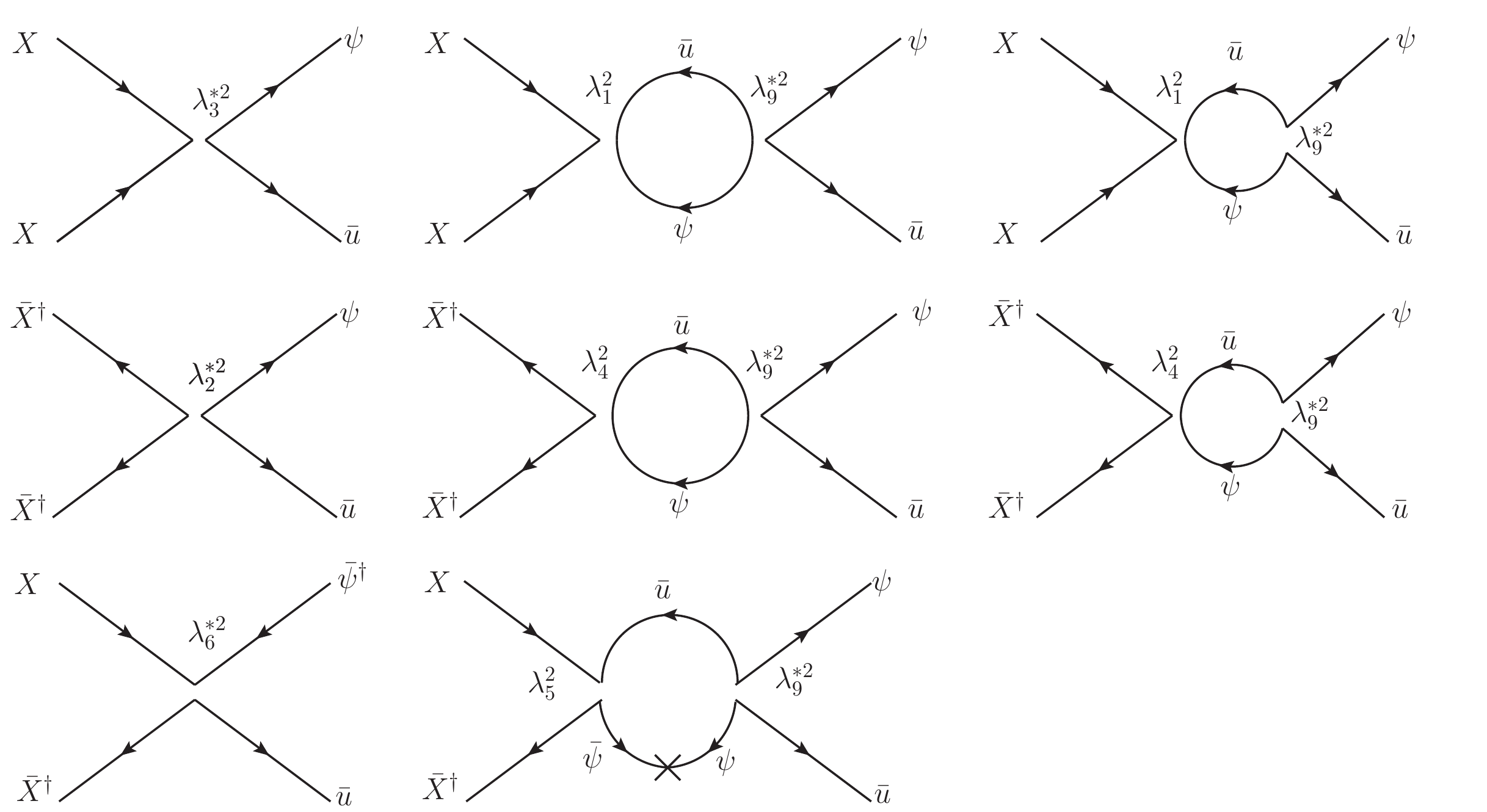}
\end{center}
\vspace{-0.8cm}
\caption{\sl \textbf{\textit{Diagrams for the annihilation process $\boldsymbol{XX \to \ubar \psi}$.}} The diagrams in the first two lines involve $s$-channel-annihilation operators, while in the last line they involve $t$-channel-annihilation operators. The cross in the last diagram represents a mass insertion, $m_\psi$. The annihilation processes $XX \to \ubardag \psidag$ are obtained from similar diagrams, with the replacements $\lambda_3 \to \lambda_1^*$, $\lambda_2 \to \lambda_4^*$, $\lambda_6 \leftrightarrow \lambda_5^*$ and $\lambda_9 \to \lambda_9^*$.}
\label{fig:asym1}
\end{figure}

The physical CP asymmetry is defined as
\be \label{eq:epsilon}
\epsilon(z) \equiv \frac{\gamma (XX\to \ubar \psi) + \gamma (\Xbar \Xbar \to \ubar \psi) -\gamma (XX \to \ubardag \psidag) - \gamma (\Xbar \Xbar \to \ubardag \psidag)  }{\gamma (XX\to \ubar \psi) + \gamma (\Xbar \Xbar \to \ubar \psi) + \gamma (XX \to \ubardag \psidag) + \gamma (\Xbar \Xbar \to \ubardag \psidag)}\,,
\ee
where the $\gamma$'s denote the thermally averaged interaction rates, as defined in the previous appendix.
Eq.~\eqref{eq:epsilon} can be calculated in our model, at leading order, from the interference between tree-level and one-loop diagrams, as shown in figure~\ref{fig:asym1}.

When we sum the squared matrix elements at the numerator and at the denominator we find the following:
\begin{eqnarray}
&&\sum \vert {\mathcal M_{\rm NUM}}\vert^2=\frac{\left(s-m_\psi^2\right)^3}{8\,\pi\,s\,\Lambda^6}\times \bigg\lbrace 2\,m_\chi^2(\mathcal{I}_{12}+\mathcal{I}_{34})+(s-2m_\chi^2)(\mathcal{I}_{13}+\mathcal{I}_{24}) \nonumber \\
&&\hspace{3cm}+ m_\psi m_\chi (\mathcal{I}_{25}-\mathcal{I}_{35}-\mathcal{I}_{16}+\frac{t-m_\chi^2}{s-m_\psi^2}(\mathcal{I}_{16}+\mathcal{I}_{46}-\frac{m_\psi}{m_\chi}\,\mathcal{I}_{56}))\,\bigg\rbrace\,, \\
&&\sum \vert {\mathcal M_{\rm DEN}}\vert^2=\frac{1}{\Lambda^4}\Big\lbrace (s-2m_\chi^2)(s-m_\psi^2)(\vert \lambda_1\vert^2+\vert \lambda_2\vert^2+\vert \lambda_3\vert^2+\vert \lambda_4\vert^2)  \nn \\
&&\hspace{3cm}+(t-m_\chi^2)(t-m_\chi^2-m_\psi^2)(\vert \lambda_5\vert^2+\vert \lambda_6\vert^2)  \nonumber \\
&&\hspace{3cm}- 4\,m_\chi^2(s-m_\psi^2)(\mathcal{R}_{14}+\mathcal{R}_{23})-2m_\psi m_\chi (t-m_\chi^2)(\mathcal{R}_{26}+\mathcal{R}_{45})\nonumber \\
&&\hspace{3cm} -2m_\psi m_\chi(s+t-m_\psi^2-m_\chi^2)(\mathcal{R}_{15}+\mathcal{R}_{36})\Big\rbrace \,,
\end{eqnarray}
with $\mathcal{I}_{ij}\equiv {\rm Im}\left(\lambda_i^2\,\lambda_j^2\,\lambda_9^{*\,2}\right)$ and $\mathcal{R}_{ij}\equiv {\rm Re}\left(\lambda_i^2\,\lambda_j^{*\,2}\right)$.
A few comments are in order at this stage. 
First, the cross sections at the denominator are computed at tree level.
Second, the combinations of the couplings that appear in the imaginary parts at the numerator are invariant under rephasing of the fields $X$, $\psi$ and $\ubar$, as one would expect for a physically meaningful result. Third, the factor of $(s-m_\psi^2)^3$ at the numerator is a consequence of the fact that the particles in the loop go on-shell.

Next we need to integrate over the phase space to obtain the cross sections, then we need to perform the integrals defined in eq.~\eqref{eq:rates}. The result we find for $\epsilon$ is quite lengthy, but, remarkably, analytic. It depends on the couplings $\lambda_{1\dots 6},\lambda_9$, on the masses $m_\psi$ and $m_\chi$, as well as on the temperature via $z = m_\chi/T$.
It is instructive to look at some limiting cases rather than at the general full expression. Setting some equalities among the couplings, as in Section~\ref{EFT},
and taking the low temperature limit ($z\to \infty$), we find 
\begin{equation}\label{epsts}
\epsilon=\frac{\vert \lambda_\text{WO}\vert ^2}{4\pi}\sin (2\delta) \left(\frac{m_\chi}{\Lambda}\right)^2\,\left(1-x^2\right)^2\frac{2 \lambda_p^4+ 3\,x\,\lambda_p^2\,\lambda_t^2+2\,x^2\lambda_t^4}{2\lambda_p^4-2\,x\lambda_p^2\,\lambda_t^2+\lambda_t^4\left(1+x^2\right)},
\end{equation}
where as before $x=  m_\psi/(2\,m_\chi)$ and 
\begin{equation}\label{lambwo}
\lambda_\text{WO}=\vert \lambda_\text{WO} \vert\, e^{i\,\delta} \equiv \lambda_9\,.
\end{equation}
With the assumed equalities among the couplings we lose the rephasing invariance of the result. This is not a big deal, given that we have shown above that  the general result is rephasing invariant, but it leads us to make a further assumption, which is to take the couplings $\lambda_p$, $\lambda_t$ real, and keep only $\lambda_\text{WO}$ complex.
This expression is the same as the one we would get by defining $\epsilon$ as a ratio of cross sections instead of thermal averages. This is no surprise, given that we have taken the zero temperature limit. We would like to remind the reader that in the literature $\epsilon$ is often defined as a ratio of widths or cross sections, instead of thermal averages. The former is an approximation which corresponds to the zero-temperature limit of the latter, and is often a good approximation~\cite{Nardi:2007jp,Fong:2010bh}.

\section{Neutron - antineutron oscillation} \label{app:nn}

\begin{figure}[t!]
\centering
\includegraphics[width=\textwidth]{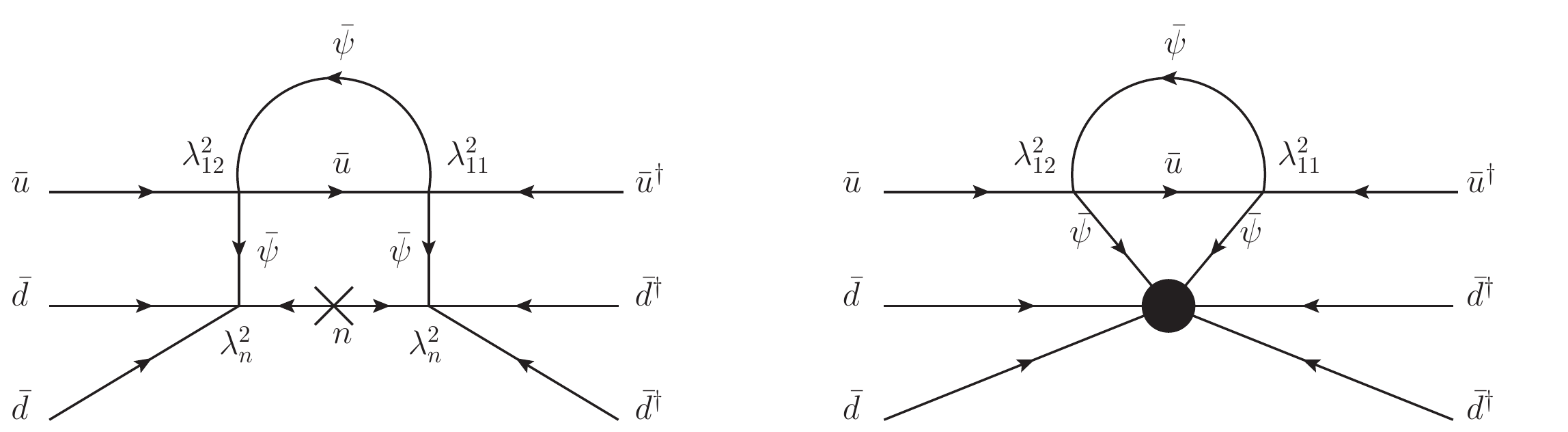}
\caption{\sl \textbf{\textit{Diagrams for $\boldsymbol{n-\bar n}$ oscillation.}} The first diagram only involves dimension six operators of our Lagrangian. The second one contains a vertex that corresponds to the dimension nine operator discussed in the text.}
\label{fig:nn}
\end{figure}

The operator $\lambda_n^2 (\psibar \bar d)(\bar d n)$ in our lagrangian violates baryon number and can contribute to neutron-antineutron mixing. The lowest order contribution comes from the two-loop diagram shown in the left of figure~\ref{fig:nn}. 
Note that the diagram contains a mass insertion for the light singlet $n$. A very rough estimate gives
\be
\delta m_\text{osc} \sim 10^{-4}\, m_n\, \frac{\lambda_{11}^2 \lambda_{12}^2 \lambda_n^4}{\Lambda^8}\,\Lambda_\text{QCD}^8 \sim 10^{-43} \ {\rm GeV}.
\ee
The factor of $10^{-4}$ comes from the two-loop suppression, $m_n$ is the singlet mass\footnote{As explained earlier on in the paper, the singlet has to be very light in order to avoid over closure of the Universe. Note that if it is massless, there is no contribution to the neutron-antineutron oscillation from the diagram under consideration.}, taken here to be $\sim$ eV, we have set the couplings $\lambda$'s to one and the scale $\Lambda$ to 1 TeV. Note this is a conservative choice, in the rest of the paper we have used $\Lambda = 10$ TeV. 
Our estimate is well below the experimental bound which can be expressed as $\delta m_\text{osc} \sim 10^{-33}$ GeV~\cite{Mohapatra:2009wp}.

One can still worry about operators of higher dimensions in our EFT expansion that can contribute to this oscillation at fewer loops, and could thus be dangerous. One such operator would be $\frac{c}{M^5}\, \ubar \bar d \bar d \ubar \bar d \bar d$. This operator of dimension nine, that contains only SM particles, would contribute at tree level. In building a UV complete theory that realizes the WIMPy baryogenesis, one needs to make sure that such an operator is not generated at a scale $M\sim$ TeV. In other words the scale $M$ here has to be different from $\Lambda$, or alternatively there should be some mechanism that results in a coupling $c \ll 1$.
The models studied in~\cite{Cui:2011ab} provide explicit examples of UV completions that are not in conflict with bounds from neutron-antineutron oscillations.

A dimension nine operator that contains the exotic quark, specific to these models, is $\frac{1}{\Lambda^5}\, \psibar \bar d \bar d \psibar \bar d \bar d$. Now we have restored $\Lambda$ as the new physics scale, because with the presence of $\psibar$ it is reasonable to expect that this operator is generated at the same scale as the others in our models. 
Then, the lowest order contribution to the oscillation comes from the second diagram in figure~\ref{fig:nn}. Taking again couplings of order one and $\Lambda\sim$ 1 TeV, one gets the estimate $\delta m_\text{osc} \sim 10^{-38}$ GeV, still safely below the limit.

\bibliographystyle{JHEP}
\bibliography{WIMPy}

\providecommand{\href}[2]{#2}\begingroup\raggedright\begin{thebibliography}{10}

\bibitem{Jungman:1995df}
G.~Jungman, M.~Kamionkowski, and K.~Griest, {\it {Supersymmetric dark matter}},
   {\em Phys. Rept.} {\bf 267} (1996) 195,
  [\href{http://xxx.lanl.gov/abs/hep-ph/9506380}{{\tt hep-ph/9506380}}].

\bibitem{Bergstrom:2000pn}
L.~Bergström, {\it {Non-baryonic dark matter: Observational evidence and
  detection methods}},  {\em Rept. Prog. Phys.} {\bf 63} (2000) 793,
  [\href{http://xxx.lanl.gov/abs/hep-ph/0002126}{{\tt hep-ph/0002126}}].

\bibitem{Munoz:2003gx}
C.~Muñoz, {\it {Dark matter detection in the light of recent experimental
  results}},  {\em Int. J. Mod. Phys.} {\bf A19} (2004) 3093,
  [\href{http://xxx.lanl.gov/abs/hep-ph/0309346}{{\tt hep-ph/0309346}}].

\bibitem{Bertone:2004pz}
G.~Bertone, D.~Hooper, and J.~Silk, {\it {Particle dark matter: Evidence,
  candidates and constraints}},  {\em Phys. Rept.} {\bf 405} (2005) 279,
  [\href{http://xxx.lanl.gov/abs/hep-ph/0404175}{{\tt hep-ph/0404175}}].

\bibitem{Bertone:2010}
G.~Bertone, ed., {\em Particle Dark Matter: Observations, Models and Searches}.
\newblock Cambridge University Press, 2010.

\bibitem{Drees:2012ji}
M.~Drees and G.~Gerbier, {\it {Mini--Review of Dark Matter: 2012}},
  \href{http://xxx.lanl.gov/abs/1204.2373}{{\tt arXiv:1204.2373}}.

\bibitem{Bergstrom:2012fi}
L.~Bergström, {\it {Dark Matter Evidence, Particle Physics Candidates and
  Detection Methods}},  \href{http://xxx.lanl.gov/abs/1205.4882}{{\tt
  arXiv:1205.4882}}.

\bibitem{Tegmark:2006az}
{\bf SDSS} Collaboration, M.~Tegmark et~al., {\it {Cosmological Constraints
  from the SDSS Luminous Red Galaxies}},  {\em Phys. Rev.} {\bf D74} (2006)
  123507, [\href{http://xxx.lanl.gov/abs/astro-ph/0608632}{{\tt
  astro-ph/0608632}}].

\bibitem{Komatsu:2010fb}
{\bf WMAP} Collaboration, E.~Komatsu et~al., {\it {Seven-Year Wilkinson
  Microwave Anisotropy Probe (WMAP) Observations: Cosmological
  Interpretation}},  {\em Astrophys. J. Suppl.} {\bf 192} (2011) 18,
  [\href{http://xxx.lanl.gov/abs/1001.4538}{{\tt arXiv:1001.4538}}].

\bibitem{Jarosik:2010iu}
{\bf WMAP} Collaboration, N.~Jarosik, C.~Bennett, J.~Dunkley, B.~Gold,
  M.~Greason, et~al., {\it {Seven-Year Wilkinson Microwave Anisotropy Probe
  (WMAP) Observations: Sky Maps, Systematic Errors, and Basic Results}},  {\em
  Astrophys.J.Suppl.} {\bf 192} (2011) 14,
  [\href{http://xxx.lanl.gov/abs/1001.4744}{{\tt arXiv:1001.4744}}].

\bibitem{Nussinov:1985xr}
S.~Nussinov, {\it {Technocosmology: could a technibaryon excess provide a
  `natural' missing mass candidate?}},  {\em Phys.Lett.} {\bf B165} (1985) 55.

\bibitem{Hooper:2004dc}
D.~Hooper, J.~March-Russell, and S.~M. West, {\it {Asymmetric sneutrino dark
  matter and the Omega(b) / Omega(DM) puzzle}},  {\em Phys.Lett.} {\bf B605}
  (2005) 228--236, [\href{http://xxx.lanl.gov/abs/hep-ph/0410114}{{\tt
  hep-ph/0410114}}].

\bibitem{Farrar:2005zd}
G.~R. Farrar and G.~Zaharijas, {\it {Dark matter and the baryon asymmetry}},
  {\em Phys.Rev.Lett.} {\bf 96} (2006) 041302,
  [\href{http://xxx.lanl.gov/abs/hep-ph/0510079}{{\tt hep-ph/0510079}}].

\bibitem{Kitano:2008tk}
R.~Kitano, H.~Murayama, and M.~Ratz, {\it {Unified origin of baryons and dark
  matter}},  {\em Phys.Lett.} {\bf B669} (2008) 145--149,
  [\href{http://xxx.lanl.gov/abs/0807.4313}{{\tt arXiv:0807.4313}}].

\bibitem{Kaplan:2009ag}
D.~E. Kaplan, M.~A. Luty, and K.~M. Zurek, {\it {Asymmetric Dark Matter}},
  {\em Phys.Rev.} {\bf D79} (2009) 115016,
  [\href{http://xxx.lanl.gov/abs/0901.4117}{{\tt arXiv:0901.4117}}].

\bibitem{Cohen:2009fz}
T.~Cohen and K.~M. Zurek, {\it {Leptophilic Dark Matter from the Lepton
  Asymmetry}},  {\em Phys.Rev.Lett.} {\bf 104} (2010) 101301,
  [\href{http://xxx.lanl.gov/abs/0909.2035}{{\tt arXiv:0909.2035}}].

\bibitem{Cai:2009ia}
Y.~Cai, M.~A. Luty, and D.~E. Kaplan, {\it {Leptonic Indirect Detection Signals
  from Strongly Interacting Asymmetric Dark Matter}},
  \href{http://xxx.lanl.gov/abs/0909.5499}{{\tt arXiv:0909.5499}}.

\bibitem{An:2009vq}
H.~An, S.-L. Chen, R.~N. Mohapatra, and Y.~Zhang, {\it {Leptogenesis as a
  Common Origin for Matter and Dark Matter}},  {\em JHEP} {\bf 1003} (2010)
  124, [\href{http://xxx.lanl.gov/abs/0911.4463}{{\tt arXiv:0911.4463}}].

\bibitem{Cohen:2010kn}
T.~Cohen, D.~J. Phalen, A.~Pierce, and K.~M. Zurek, {\it {Asymmetric Dark
  Matter from a GeV Hidden Sector}},  {\em Phys.Rev.} {\bf D82} (2010) 056001,
  [\href{http://xxx.lanl.gov/abs/1005.1655}{{\tt arXiv:1005.1655}}].

\bibitem{Shelton:2010ta}
J.~Shelton and K.~M. Zurek, {\it {Darkogenesis: A baryon asymmetry from the
  dark matter sector}},  {\em Phys.Rev.} {\bf D82} (2010) 123512,
  [\href{http://xxx.lanl.gov/abs/1008.1997}{{\tt arXiv:1008.1997}}].

\bibitem{Davoudiasl:2010am}
H.~Davoudiasl, D.~E. Morrissey, K.~Sigurdson, and S.~Tulin, {\it {Hylogenesis:
  A Unified Origin for Baryonic Visible Matter and Antibaryonic Dark Matter}},
  {\em Phys.Rev.Lett.} {\bf 105} (2010) 211304,
  [\href{http://xxx.lanl.gov/abs/1008.2399}{{\tt arXiv:1008.2399}}].

\bibitem{Haba:2010bm}
N.~Haba and S.~Matsumoto, {\it {Baryogenesis from Dark Sector}},  {\em
  Prog.Theor.Phys.} {\bf 125} (2011) 1311--1316,
  [\href{http://xxx.lanl.gov/abs/1008.2487}{{\tt arXiv:1008.2487}}].

\bibitem{Chun:2010hz}
E.~J. Chun, {\it {Leptogenesis origin of Dirac gaugino dark matter}},  {\em
  Phys.Rev.} {\bf D83} (2011) 053004,
  [\href{http://xxx.lanl.gov/abs/1009.0983}{{\tt arXiv:1009.0983}}].

\bibitem{Gu:2010ft}
P.-H. Gu, M.~Lindner, U.~Sarkar, and X.~Zhang, {\it {WIMP Dark Matter and
  Baryogenesis}},  {\em Phys.Rev.} {\bf D83} (2011) 055008,
  [\href{http://xxx.lanl.gov/abs/1009.2690}{{\tt arXiv:1009.2690}}].

\bibitem{Blennow:2010qp}
M.~Blennow, B.~Dasgupta, E.~Fern\'andez-Mart\'inez, and N.~Rius, {\it
  {Aidnogenesis via Leptogenesis and Dark Sphalerons}},  {\em JHEP} {\bf 1103}
  (2011) 014, [\href{http://xxx.lanl.gov/abs/1009.3159}{{\tt
  arXiv:1009.3159}}].

\bibitem{McDonald:2011zz}
J.~McDonald, {\it {Baryomorphosis: Relating the Baryon Asymmetry to the `WIMP
  Miracle'}},  {\em Phys.Rev.} {\bf D83} (2011) 083509,
  [\href{http://xxx.lanl.gov/abs/1009.3227}{{\tt arXiv:1009.3227}}].

\bibitem{Allahverdi:2010rh}
R.~Allahverdi, B.~Dutta, and K.~Sinha, {\it {Cladogenesis: Baryon-Dark Matter
  Coincidence from Branchings in Moduli Decay}},  {\em Phys.Rev.} {\bf D83}
  (2011) 083502, [\href{http://xxx.lanl.gov/abs/1011.1286}{{\tt
  arXiv:1011.1286}}].

\bibitem{Dutta:2010va}
B.~Dutta and J.~Kumar, {\it {Asymmetric Dark Matter from Hidden Sector
  Baryogenesis}},  {\em Phys.Lett.} {\bf B699} (2011) 364--367,
  [\href{http://xxx.lanl.gov/abs/1012.1341}{{\tt arXiv:1012.1341}}].

\bibitem{Falkowski:2011xh}
A.~Falkowski, J.~T. Ruderman, and T.~Volansky, {\it {Asymmetric Dark Matter
  from Leptogenesis}},  {\em JHEP} {\bf 1105} (2011) 106,
  [\href{http://xxx.lanl.gov/abs/1101.4936}{{\tt arXiv:1101.4936}}].

\bibitem{Heckman:2011sw}
J.~J. Heckman and S.-J. Rey, {\it {Baryon and Dark Matter Genesis from Strongly
  Coupled Strings}},  {\em JHEP} {\bf 1106} (2011) 120,
  [\href{http://xxx.lanl.gov/abs/1102.5346}{{\tt arXiv:1102.5346}}].

\bibitem{Frandsen:2011kt}
M.~T. Frandsen, S.~Sarkar, and K.~Schmidt-Hoberg, {\it {Light asymmetric dark
  matter from new strong dynamics}},  {\em Phys.Rev.} {\bf D84} (2011) 051703,
  [\href{http://xxx.lanl.gov/abs/1103.4350}{{\tt arXiv:1103.4350}}].

\bibitem{Buckley:2011kk}
M.~R. Buckley, {\it {Asymmetric Dark Matter and Effective Operators}},  {\em
  Phys.Rev.} {\bf D84} (2011) 043510,
  [\href{http://xxx.lanl.gov/abs/1104.1429}{{\tt arXiv:1104.1429}}].

\bibitem{Iminniyaz:2011yp}
H.~Iminniyaz, M.~Drees, and X.~Chen, {\it {Relic Abundance of Asymmetric Dark
  Matter}},  {\em JCAP} {\bf 1107} (2011) 003,
  [\href{http://xxx.lanl.gov/abs/1104.5548}{{\tt arXiv:1104.5548}}].

\bibitem{Cheung:2011if}
C.~Cheung and K.~M. Zurek, {\it {Affleck-Dine Cogenesis}},  {\em Phys.Rev.}
  {\bf D84} (2011) 035007, [\href{http://xxx.lanl.gov/abs/1105.4612}{{\tt
  arXiv:1105.4612}}].

\bibitem{MarchRussell:2011fi}
J.~March-Russell and M.~McCullough, {\it {Asymmetric Dark Matter via
  Spontaneous Co-Genesis}},  {\em JCAP} {\bf 1203} (2012) 019,
  [\href{http://xxx.lanl.gov/abs/1106.4319}{{\tt arXiv:1106.4319}}].

\bibitem{Davoudiasl:2011fj}
H.~Davoudiasl, D.~E. Morrissey, K.~Sigurdson, and S.~Tulin, {\it {Baryon
  Destruction by Asymmetric Dark Matter}},  {\em Phys.Rev.} {\bf D84} (2011)
  096008, [\href{http://xxx.lanl.gov/abs/1106.4320}{{\tt arXiv:1106.4320}}].

\bibitem{Cui:2011qe}
Y.~Cui, L.~Randall, and B.~Shuve, {\it {Emergent Dark Matter, Baryon, and
  Lepton Numbers}},  {\em JHEP} {\bf 1108} (2011) 073,
  [\href{http://xxx.lanl.gov/abs/1106.4834}{{\tt arXiv:1106.4834}}].

\bibitem{Arina:2011cu}
C.~Arina and N.~Sahu, {\it {Asymmetric Inelastic Inert Doublet Dark Matter from
  Triplet Scalar Leptogenesis}},  {\em Nucl.Phys.} {\bf B854} (2012) 666--699,
  [\href{http://xxx.lanl.gov/abs/1108.3967}{{\tt arXiv:1108.3967}}].

\bibitem{McDonald:2011sv}
J.~McDonald, {\it {Simultaneous Generation of WIMP Miracle-like Densities of
  Baryons and Dark Matter}},  {\em Phys.Rev.} {\bf D84} (2011) 103514,
  [\href{http://xxx.lanl.gov/abs/1108.4653}{{\tt arXiv:1108.4653}}].

\bibitem{Cirelli:2011ac}
M.~Cirelli, P.~Panci, G.~Servant, and G.~Zaharijas, {\it {Consequences of
  DM/antiDM Oscillations for Asymmetric WIMP Dark Matter}},  {\em JCAP} {\bf
  1203} (2012) 015, [\href{http://xxx.lanl.gov/abs/1110.3809}{{\tt
  arXiv:1110.3809}}].

\bibitem{Chowdhury:2011ga}
T.~A. Chowdhury, M.~Nemevsek, G.~Senjanovic, and Y.~Zhang, {\it {Dark Matter as
  the Trigger of Strong Electroweak Phase Transition}},  {\em JCAP} {\bf 1202}
  (2012) 029, [\href{http://xxx.lanl.gov/abs/1110.5334}{{\tt
  arXiv:1110.5334}}].

\bibitem{Kamada:2012ht}
K.~Kamada and M.~Yamaguchi, {\it {Asymmetric Dark Matter from Spontaneous
  Cogenesis in the Supersymmetric Standard Model}},  {\em Phys.Rev.} {\bf D85}
  (2012) 103530, [\href{http://xxx.lanl.gov/abs/1201.2636}{{\tt
  arXiv:1201.2636}}].

\bibitem{Blum:2012nf}
K.~Blum, A.~Efrati, Y.~Grossman, Y.~Nir, and A.~Riotto, {\it {Asymmetric
  Higgsino Dark Matter}},  {\em Phys.Rev.Lett.} {\bf 109} (2012) 051302,
  [\href{http://xxx.lanl.gov/abs/1201.2699}{{\tt arXiv:1201.2699}}].

\bibitem{Tulin:2012re}
S.~Tulin, H.-B. Yu, and K.~M. Zurek, {\it {Oscillating Asymmetric Dark
  Matter}},  {\em JCAP} {\bf 1205} (2012) 013,
  [\href{http://xxx.lanl.gov/abs/1202.0283}{{\tt arXiv:1202.0283}}].

\bibitem{Walker:2012ka}
D.~G. Walker, {\it {Dark Baryogenesis}},
  \href{http://xxx.lanl.gov/abs/1202.2348}{{\tt arXiv:1202.2348}}.

\bibitem{Davoudiasl:2012uw}
H.~Davoudiasl and R.~N. Mohapatra, {\it {On Relating the Genesis of Cosmic
  Baryons and Dark Matter}},  {\em New J.Phys.} {\bf 14} (2012) 095011,
  [\href{http://xxx.lanl.gov/abs/1203.1247}{{\tt arXiv:1203.1247}}].

\bibitem{MarchRussell:2012hi}
J.~March-Russell, J.~Unwin, and S.~M. West, {\it {Closing in on Asymmetric Dark
  Matter I: Model independent limits for interactions with quarks}},  {\em
  JHEP} {\bf 1208} (2012) 029, [\href{http://xxx.lanl.gov/abs/1203.4854}{{\tt
  arXiv:1203.4854}}].

\bibitem{Feng:2012jn}
W.-Z. Feng, P.~Nath, and G.~Peim, {\it {Cosmic Coincidence and Asymmetric Dark
  Matter in a Stueckelberg Extension}},  {\em Phys.Rev.} {\bf D85} (2012)
  115016, [\href{http://xxx.lanl.gov/abs/1204.5752}{{\tt arXiv:1204.5752}}].

\bibitem{Ellwanger:2012yg}
U.~Ellwanger and P.~Mitropoulos, {\it {Upper Bounds on Asymmetric Dark Matter
  Self Annihilation Cross Sections}},  {\em JCAP} {\bf 1207} (2012) 024,
  [\href{http://xxx.lanl.gov/abs/1205.0673}{{\tt arXiv:1205.0673}}].

\bibitem{Okada:2012rm}
N.~Okada and O.~Seto, {\it {Originally Asymmetric Dark Matter}},  {\em
  Phys.Rev.} {\bf D86} (2012) 063525,
  [\href{http://xxx.lanl.gov/abs/1205.2844}{{\tt arXiv:1205.2844}}].

\bibitem{Gu:2012fg}
P.-H. Gu, {\it {From Dirac neutrino masses to baryonic and dark matter
  asymmetries}},  \href{http://xxx.lanl.gov/abs/1209.4579}{{\tt
  arXiv:1209.4579}}.

\bibitem{D'Eramo:2011ec}
F.~D'Eramo, L.~Fei, and J.~Thaler, {\it {Dark Matter Assimilation into the
  Baryon Asymmetry}},  {\em JCAP} {\bf 1203} (2012) 010,
  [\href{http://xxx.lanl.gov/abs/1111.5615}{{\tt arXiv:1111.5615}}].

\bibitem{Cui:2011ab}
Y.~Cui, L.~Randall, and B.~Shuve, {\it {A WIMPy Baryogenesis Miracle}},  {\em
  JHEP} {\bf 1204} (2012) 075, [\href{http://xxx.lanl.gov/abs/1112.2704}{{\tt
  arXiv:1112.2704}}].

\bibitem{Dreiner:2008tw}
H.~K. Dreiner, H.~E. Haber, and S.~P. Martin, {\it {Two-component spinor
  techniques and Feynman rules for quantum field theory and supersymmetry}},
  {\em Phys.Rept.} {\bf 494} (2010) 1--196,
  [\href{http://xxx.lanl.gov/abs/0812.1594}{{\tt arXiv:0812.1594}}].

\bibitem{Cao:2009uw}
Q.-H. Cao, C.-R. Chen, C.~S. Li, and H.~Zhang, {\it {Effective Dark Matter
  Model: Relic density, CDMS II, Fermi LAT and LHC}},  {\em JHEP} {\bf 1108}
  (2011) 018, [\href{http://xxx.lanl.gov/abs/0912.4511}{{\tt
  arXiv:0912.4511}}].

\bibitem{Zheng:2010js}
J.-M. Zheng, Z.-H. Yu, J.-W. Shao, X.-J. Bi, Z.~Li, et~al., {\it {Constraining
  the interaction strength between dark matter and visible matter: I. fermionic
  dark matter}},  {\em Nucl.Phys.} {\bf B854} (2012) 350--374,
  [\href{http://xxx.lanl.gov/abs/1012.2022}{{\tt arXiv:1012.2022}}].

\bibitem{Cheung:2012gi}
K.~Cheung, P.-Y. Tseng, Y.-L.~S. Tsai, and T.-C. Yuan, {\it {Global Constraints
  on Effective Dark Matter Interactions: Relic Density, Direct Detection,
  Indirect Detection, and Collider}},  {\em JCAP} {\bf 1205} (2012) 001,
  [\href{http://xxx.lanl.gov/abs/1201.3402}{{\tt arXiv:1201.3402}}].

\bibitem{Sakharov:1967dj}
A.~Sakharov, {\it {Violation of CP Invariance, C Asymmetry, and Baryon
  Asymmetry of the Universe}},  {\em Pisma Zh.Eksp.Teor.Fiz.} {\bf 5} (1967)
  32--35.

\bibitem{CMS-PAS-SUS-11-016}
{\bf CMS} Collaboration, ``{\it Interpretation of Searches for
  Supersymmetry}.'' \url{http://cdsweb.cern.ch/record/1445580?ln=en}, 2012.

\bibitem{ATLAS}
{\bf ATLAS} Collaboration, G.~Aad et~al., {\it {Search for squarks and gluinos
  with the ATLAS detector in final states with jets and missing transverse
  momentum using 4.7 fb$^{-1}$ of $sqrt(s) = 7$ TeV proton-proton collision
  data}},  \href{http://xxx.lanl.gov/abs/1208.0949}{{\tt arXiv:1208.0949}}.

\bibitem{Aprile:2012nq}
{\bf XENON100 Collaboration} Collaboration, E.~Aprile et~al., {\it {Dark Matter
  Results from 225 Live Days of XENON100 Data}},  {\em Phys.Rev.Lett.} {\bf
  109} (2012) 181301, [\href{http://xxx.lanl.gov/abs/1207.5988}{{\tt
  arXiv:1207.5988}}].

\bibitem{Akimov:2011tj}
D.~Y. Akimov, H.~Araujo, E.~Barnes, V.~Belov, A.~Bewick, et~al., {\it
  {WIMP-nucleon cross-section results from the second science run of
  ZEPLIN-III}},  {\em Phys.Lett.} {\bf B709} (2012) 14--20,
  [\href{http://xxx.lanl.gov/abs/1110.4769}{{\tt arXiv:1110.4769}}].

\bibitem{Felizardo:2011uw}
M.~Felizardo, T.~Girard, T.~Morlat, A.~Fernandes, A.~Ramos, et~al., {\it {Final
  Analysis and Results of the Phase II SIMPLE Dark Matter Search}},  {\em
  Phys.Rev.Lett.} {\bf 108} (2012) 201302,
  [\href{http://xxx.lanl.gov/abs/1106.3014}{{\tt arXiv:1106.3014}}].

\bibitem{Bai:2012xg}
Y.~Bai and T.~M. Tait, {\it {Searches with Mono-Leptons}},
  \href{http://xxx.lanl.gov/abs/1208.4361}{{\tt arXiv:1208.4361}}.

\bibitem{Goodman:2010ku}
J.~Goodman, M.~Ibe, A.~Rajaraman, W.~Shepherd, T.~M. Tait, et~al., {\it
  {Constraints on Dark Matter from Colliders}},  {\em Phys.Rev.} {\bf D82}
  (2010) 116010, [\href{http://xxx.lanl.gov/abs/1008.1783}{{\tt
  arXiv:1008.1783}}].

\bibitem{Fox:2011pm}
P.~J. Fox, R.~Harnik, J.~Kopp, and Y.~Tsai, {\it {Missing Energy Signatures of
  Dark Matter at the LHC}},  {\em Phys.Rev.} {\bf D85} (2012) 056011,
  [\href{http://xxx.lanl.gov/abs/1109.4398}{{\tt arXiv:1109.4398}}].

\bibitem{PhysRevD.86.010001}
{\bf Particle Data Group} Collaboration, J.~Beringer et~al., {\it Review of
  particle physics},  {\em Phys. Rev. D} {\bf 86} (Jul, 2012) 010001.

\bibitem{Kawasaki:2004qu}
M.~Kawasaki, K.~Kohri, and T.~Moroi, {\it {Big-Bang nucleosynthesis and
  hadronic decay of long-lived massive particles}},  {\em Phys.Rev.} {\bf D71}
  (2005) 083502, [\href{http://xxx.lanl.gov/abs/astro-ph/0408426}{{\tt
  astro-ph/0408426}}].

\bibitem{Nardi:2007jp}
E.~Nardi, J.~Racker, and E.~Roulet, {\it {CP violation in scatterings, three
  body processes and the Boltzmann equations for leptogenesis}},  {\em JHEP}
  {\bf 0709} (2007) 090, [\href{http://xxx.lanl.gov/abs/0707.0378}{{\tt
  arXiv:0707.0378}}].

\bibitem{Fong:2010bh}
C.~S. Fong, M.~González-García, and J.~Racker, {\it {CP Violation from
  Scatterings with Gauge Bosons in Leptogenesis}},  {\em Phys.Lett.} {\bf B697}
  (2011) 463--470, [\href{http://xxx.lanl.gov/abs/1010.2209}{{\tt
  arXiv:1010.2209}}].

\bibitem{Mohapatra:2009wp}
R.~Mohapatra, {\it {Neutron-Anti-Neutron Oscillation: Theory and
  Phenomenology}},  {\em J.Phys.} {\bf G36} (2009) 104006,
  [\href{http://xxx.lanl.gov/abs/0902.0834}{{\tt arXiv:0902.0834}}].

\end{thebibliography}\endgroup


30 atime=1359124318.276808376
30 ctime=1355929897.963216492

\end{document}